\documentclass[twocolumn,preprint]{aastex63}
\usepackage{amsmath,amstext}
\usepackage[T1]{fontenc}
\usepackage{apjfonts} 
\usepackage{natbib}
\usepackage{longtable}
\usepackage{comment}

\newcommand{\hst}{\textit{HST}}
\newcommand{\hstlong}{\textit{Hubble Space Telescope}}
\newcommand{\jwst}{\textit{JWST}}
\newcommand{\jwstlong}{\textit{James Webb Space Telescope}}
\newcommand{\romanlong}{\textit{Nancy Grace Roman Space Telescope}}

\newcommand{\grizli}{\texttt{grizli}}
\newcommand{\pyneb}{\texttt{PyNeb}}

\newcommand{\Ha}{\hbox{{\rm H}$\alpha$}}

\newcommand{\Hb}{\hbox{{\rm H}$\beta$}}
\newcommand{\hb}{\hbox{{\rm H}$\beta$}}

\newcommand{\neiii}{\hbox{[\ion{Ne}{3}]}}
\newcommand{\nev}{\hbox{[\ion{Ne}{5}]}}
\newcommand{\oii}{\hbox{[\ion{O}{2}]}}
\newcommand{\oiii}{\hbox{[\ion{O}{3}]}}

\newcommand{\sii}{\hbox{[\ion{S}{2}]}}
\newcommand{\nii}{\hbox{[\ion{N}{2}]}}

\newcommand{\zvo}{\hbox{$1.39 < z < 1.45$}}
\newcommand{\zohno}{\hbox{$1.39 < z < 2.30$}}
\newcommand{\snr}{\hbox{S/N}}

\begin{document}

\title{\large \bf CLEAR: High-Ionization \nev\ $\mathbf{\lambda 3426}$ Emission-line Galaxies at $\mathbf{1.4<}z\mathbf{<2.3}$}

\correspondingauthor{Nikko J. Cleri}
\email{cleri@tamu.edu}

\author[0000-0001-7151-009X]{Nikko J. Cleri}
\affiliation{Department of Physics and Astronomy, Texas A\&M University, College Station, TX, 77843-4242 USA}
\affiliation{George P.\ and Cynthia Woods Mitchell Institute for Fundamental Physics and Astronomy, Texas A\&M University, College Station, TX, 77843-4242 USA}
\affiliation{Department of Physics, University of Connecticut, Storrs, CT 06269, USA}

\author[0000-0001-8835-7722]{Guang Yang}
\affiliation{Kapteyn Astronomical Institute, University of Groningen, P.O. Box 800, 9700 AV Groningen, The Netherlands}
\affiliation{SRON Netherlands Institute for Space Research, Postbus 800, 9700 AV Groningen, The Netherlands}
\affiliation{Department of Physics and Astronomy, Texas A\&M University, College Station, TX, 77843-4242 USA}
\affiliation{George P.\ and Cynthia Woods Mitchell Institute for Fundamental Physics and Astronomy, Texas A\&M University, College Station, TX, 77843-4242 USA}

\author[0000-0001-7503-8482]{Casey Papovich}
\affiliation{Department of Physics and Astronomy, Texas A\&M University, College Station, TX, 77843-4242 USA}
\affiliation{George P.\ and Cynthia Woods Mitchell Institute for Fundamental Physics and Astronomy, Texas A\&M University, College Station, TX, 77843-4242 USA}

\author[0000-0002-1410-0470]{Jonathan R. Trump}
\affiliation{Department of Physics, University of Connecticut, Storrs, CT 06269, USA}

\author[0000-0001-8534-7502]{Bren E. Backhaus}
\affiliation{Department of Physics, University of Connecticut, Storrs, CT 06269, USA}

\author[0000-0001-8489-2349]{Vicente Estrada-Carpenter}
\affiliation{Department of Astronomy \& Physics, Saint Mary's University, 923 Robie Street, Halifax, NS, B3H 3C3, Canada}
\affiliation{Department of Physics and Astronomy, Texas A\&M University, College Station, TX, 77843-4242 USA}
\affiliation{George P.\ and Cynthia Woods Mitchell Institute for Fundamental Physics and Astronomy, Texas A\&M University, College Station, TX, 77843-4242 USA}

\author[0000-0001-8519-1130]{Steven L. Finkelstein}
\affiliation{Department of Astronomy, The University of Texas, Austin, Texas, 78712 USA} 

\author[0000-0002-7831-8751]{Mauro Giavalisco}
\affiliation{Department of Astronomy, University of Massachusetts Amherst, 710 N. Pleasant St., Amherst, MA, 01003, USA}

\author[0000-0001-6251-4988]{Taylor A. Hutchison}
\altaffiliation{NASA Postdoctoral Fellow}
\affiliation{Astrophysics Science Division, NASA Goddard Space Flight Center, 8800 Greenbelt Rd, Greenbelt, MD 20771, USA}
\affiliation{Department of Physics and Astronomy, Texas A\&M University, College Station, TX, 77843-4242 USA}
\affiliation{George P.\ and Cynthia Woods Mitchell Institute for Fundamental Physics and Astronomy, Texas A\&M University, College Station, TX, 77843-4242 USA}

\author[0000-0001-7673-2257]{Zhiyuan Ji}
\affiliation{Department of Astronomy, University of Massachusetts Amherst, 710 N. Pleasant St., Amherst, MA, 01003, USA}

\author[0000-0003-1187-4240]{Intae Jung}
\affiliation{Department of Physics, The Catholic University of America, Washington, DC 20064, USA}
\affiliation{Astrophysics Science Division, Goddard Space Flight Center, Greenbelt, MD 20771, USA}
\affiliation{Center for Research and Exploration in Space Science and Technology, NASA/GSFC, Greenbelt, MD 20771}

\author[0000-0002-7547-3385]{Jasleen Matharu}
\affiliation{Cosmic Dawn Center, Niels Bohr Institute, University of Copenhagen, R\aa dmandsgade 62, 2200 Copenhagen, Denmark\\}
\affiliation{Department of Physics and Astronomy, Texas A\&M University, College Station, TX, 77843-4242 USA}
\affiliation{George P.\ and Cynthia Woods Mitchell Institute for Fundamental Physics and Astronomy, Texas A\&M University, College Station, TX, 77843-4242 USA}

\author[0000-0003-1665-2073]{Ivelina Momcheva}
\affiliation{Max-Planck-Institut für Astronomie, Königstuhl 17, D-69117 Heidelberg, Germany}
\affiliation{Space Telescope Science Institute, 3700 San Martin Drive, Baltimore, MD, 21218 USA}

\author[0000-0002-4606-4240]{Grace M. Olivier}
\affiliation{Department of Physics and Astronomy, Texas A\&M University, College Station, TX, 77843-4242 USA}
\affiliation{George P.\ and Cynthia Woods Mitchell Institute for Fundamental Physics and Astronomy, Texas A\&M University, College Station, TX, 77843-4242 USA}

\author[0000-0002-6386-7299]{Raymond Simons}
\affiliation{Space Telescope Science Institute, 3700 San Martin Drive, Baltimore, MD, 21218 USA}

\author[0000-0001-6065-7483]{Benjamin Weiner}
\affiliation{MMT/Steward Observatory, 933 N. Cherry St., University of Arizona, Tucson, AZ 85721, USA}

\begin{abstract}
We analyze a sample of 25 \nev\ $\lambda$3426 emission-line galaxies at $1.4<z<2.3$ using \hstlong/\textit{Wide Field Camera 3} G102 and G141 grism observations from the CANDELS Lyman-$\alpha$ Emission at Reionization (CLEAR) survey. \nev\ emission probes extremely energetic photoionization (97.11-126.21 eV), and is often attributed to energetic radiation from active galactic nuclei (AGN), shocks from supernova, or an otherwise very hard ionizing spectrum from the stellar continuum. In this work, we use \nev\ in conjunction with other rest-frame UV/optical emission lines (\oii\ $\lambda\lambda$3726,3729, \neiii\ $\lambda$3869, \hb, \oiii\ $\lambda\lambda$4959,5007, \Ha+\nii\ $\lambda\lambda$6548,6583, \sii\ $\lambda\lambda$6716,6731), deep (2--7 Ms) X-ray observations (from \textit{Chandra}), and mid-infrared imaging (from \textit{Spitzer}) to study the origin of this emission and to place constraints on the nature of the ionizing engine. The majority of the \nev-detected galaxies have properties consistent with ionization from AGN.  However, for our \nev-selected sample, the X-ray luminosities are consistent with local ($z\lesssim 0.1$) X-ray-selected Seyferts, but the \nev\ luminosities are more consistent with those from $z\sim 1$ X-ray-selected QSOs. The excess \nev\ emission requires either reduced hard X-rays, or a $\sim$0.1 keV excess. We discuss possible origins of the apparent \nev\ excess, which could be related to the ``soft (X-ray) excess'' observed in some QSOs and Seyferts, and/or be a consequence of a complex/anisotropic geometry for the narrow line region, combined with absorption from a warm, relativistic wind ejected from the accretion disk. We also consider implications for future studies of extreme high-ionization systems in the epoch of reionization ($z \gtrsim 6$) with the \jwstlong. 
\end{abstract}

\section{Introduction} \label{sec:intro}
Studies from modern observatories like the \hstlong\ (\hst) have shown that cosmic star-formation density peaked roughly 7-11 billion years ago ($z{\sim}1-3$), an epoch when the properties of galaxies are fundamentally different than at present \citep{Madau2014}.  Understanding the physical conditions in galaxies at these redshifts and beyond is paramount, especially as we move into the era of the \jwstlong\ (\jwst), where we expect to see increasing numbers of ``chemically young'' (i.e., low-metallicity) galaxies that are likely key contributors to the reionization of the Universe. These rapidly star-forming galaxies exhibit prominent high-ionization nebular emission-lines in their rest frame ultraviolet (UV) and optical spectra, suggesting that these reionization era galaxies are characterized by extreme radiation fields \citep[e.g.,][]{Trump2023,Katz2023,Brinchmann2022}. 

The underlying physics of these high-ionization systems of the epoch of reionization (EoR, $z>6$) remain poorly understood, and much of the information about the EoR is extrapolated from local metal-poor dwarf galaxies \citep[e.g.,][]{Olivier2022,Berg2019,Berg2021}. One means to test the ionizing sources in galaxies across cosmic time is through ratios of strong optical and near-UV emission lines, which serve as useful diagnostics of conditions in the interstellar medium (ISM) \citep[e.g.,][]{Kewley2019b}. Much of the knowledge of the physics of higher-redshift star-formation is derived from bright Balmer lines of hydrogen (\Ha\ and \Hb), along with lines of oxygen (\oii\ $\lambda\lambda$3726,3728 and \oiii\ $\lambda\lambda$4959, 5007), sulfur (\sii\ $\lambda\lambda$6717,6731) and nitrogen (\nii\ $\lambda 6584$). This suite of near-UV/optical emission lines used for spectral classifications of galaxies is optimized for observation by \hst\ around the peak of cosmic star formation at $z\sim 2$, where these lines are redshifted into the near-IR.

Previous work using this suite of emission features in the optical and near-UV have shown that, for galaxies around cosmic noon ($z\sim 2$), oxygen abundances are lower, ionization parameters are higher, and ionization fields are harder at similar stellar masses than $z\sim 0$ galaxies \citep[e.g.,][]{Erb2006,Nakajima2014,Steidel2014,Tang2021}. 

Other bright near-UV/optical emission lines remain largely unexplored. The high ionization lines of neon have been only studied slightly, focusing on the \neiii $\lambda3869$ (40.96-63.45 eV) line \citep[e.g.,][]{Masters2014,Levesque2014,Zeimann2015,Backhaus2022a}. These works have shown that \neiii\ traces \oiii\ emission, and that the \neiii/\oii\ ratio can be used as a spectral classifier of AGN and star formation in conjunction with \oiii/\Hb. 

In this work, we study an even higher energy near-UV emission feature: quadruply-ionized neon (\nev\ $\lambda\lambda$3346,3426). The energy required to produce \nev\ photons (97.11-126.21 eV) is extremely high compared to other bright UV/optical emission lines; the minimum bound is nearly triple that of \oiii\ (35.12 eV) and nearly double that of ionized helium, \ion{He}{2} (54.42 eV), which denotes the boundary of ``high ionization'' and ``very high ionization'' in the four-zone ionization model of \cite{Berg2021}.

The production of such a high ionization emission line requires an extreme photoionizing source. Studies attribute \nev\ production to photoionization from active galactic nuclei (AGN), stellar light from the extremely hot ionizing spectra (e.g., Wolf-Rayet stars), or energetic shocks from supernovae \citep{Zeimann2015,Backhaus2022a,Gilli2010,Izotov2012,Mignoli2013}. 

Studies of local star-forming galaxies have  attempted to explain the  \nev\ production through energetic supernova shocks. \cite{Izotov2012} finds five oxygen-poor blue compact dwarf (BCDs) galaxies with \nev\ emission which have \nev/\ion{He}{2} flux ratios reproducible by radiative shock models with shock velocities in the 300-500 km s$^{-1}$ range and shock ionizing contributions $\sim10\%$ that of stellar continuum ionization. However, this modeling cannot conclusively discount this $\sim10\%$ contribution, responsible for \nev\ emission, from being produced by AGN. These studies have primarily focused on low-mass galaxies (BCDs in the case of \cite{Izotov2012}). However, there are other examples, including \cite{Leung2021}, which studied extended \nev\ emission in the local ultra-luminous infrared galaxy MRK273 and showed that \nev\ is consistent with production from shocks for this object.  Therefore, shocked gas is also a viable mechanism for \nev\ emission in and around galaxies.

Emission from \nev\ has been used to study conditions in the narrow-line region (NLR) in AGN. \cite{Gilli2010} and \cite{Mignoli2013} used \nev\ luminosities in conjunction with hard-band (2-10 keV) X-ray luminosities to probe highly obscured/Compton Thick (CT) AGN. These analyses with \nev\ luminosities are inspired by similar analyses using X-ray and \oiii\ luminosities \citep{Maiolino1998,Yan2011,Lambrides2020,Heckman2005}, with the added benefit of the extreme energies required to produce \nev. \cite{Gilli2010} and \cite{Mignoli2013} conclude that galaxies with very low (<15) X-ray/\nev\ luminosity ratios are effectively all CT AGN, though the relation of the absorption column densities ($N_H$) from X-ray spectral fits to the X-ray/\nev\ ratio is highly dependent on model assumptions \citep{Li2019}.  Therefore, while low X-ray--to-\nev\ ratios indicate CT AGN in Seyferts and higher-redshift QSOs, such data have not been extended to more modest galaxies (including the the galaxies that dominate the cosmic star-formation rate density \citep{Madau2014} and SMBH accretion \citep{Hickox2018}).

In this work, we study the properties of galaxies with \nev\ at redshifts $1.4 < z < 2.3$.  We combine the \nev\ emission with information from several of the other bright rest frame UV/optical emission-line features of \neiii, \oiii, \oii, \Hb, \Ha, and \sii, to study the physical characteristics of highly-ionizing radiation in galaxies around the peak of cosmic star formation at $z\sim 2$.  Because these lines trace a range of ionization state, these lines offer direct traces of multiple phases in the ISM  \citep{Berg2021}.  By studying multiple transitions within the same (and multiple) elements, we can more clearly understand the chemical characteristics of a galaxy.

Understanding this population of high-ionization galaxies has important implications for studies of the EoR with \jwst\ \citep[e.g.,]{Trump2023,Katz2023,Rhoads2022}. The Near-IR Camera (NIRCam) and Near-IR Spectrograph (NIRSpec) on \jwst\ have the wavelength coverage to detect these extreme \nev-emitting galaxies at $0.8\lesssim z\lesssim14$, and the Near-IR Imager and Slitless Spectrograph (NIRISS) will give slitless spectroscopy of \nev\ at lower redshifts ($3<z<7$). These objects are sure to be critical targets in spectroscopic surveys with \jwst\ in the near future.

The remainder of this work is as follows: Section \ref{sec:data} describes our parent sample from the CLEAR survey and our selection of \nev\ sources. Section \ref{sec:sf_agn} compares UV/optical emission-line ratios as diagnostics of AGN activity. Section \ref{sec:nev_x} explores the the X-ray/\nev\ ratio along with X-ray and \nev\ luminosity functions. Section \ref{sec:discussion} discusses the implications of our results. Section \ref{sec:conclusions} summarizes the results of this work and discusses future studies of high-ionization galaxies with \jwst\ and the \romanlong. 

Throughout this work, we assume a flat $\Lambda$CDM cosmology with $H_0 = 70$ km s$^{-1}$ Mpc$^{-1}$ and $\Omega_M = 0.30$ \citep{Planck2020}. 

\section{Data} \label{sec:data}
Our data come from the CANDELS Lyman-$\alpha$ Emission at Reionization (CLEAR)\footnote{\texttt{https://clear.physics.tamu.edu}} survey (a Cycle 23 $\hst$ program, PI: Papovich; \citealt{Simons2023}), which consists of deep (12-orbit depth) \hst/WFC3 G102 slitless grism spectroscopy covering 0.8-1.15~\micron\ within 12 fields split between the GOODS-North (GN) and GOODS-South (GS) extragalactic survey fields (\citealt{Estrada-Carpenter2019,Simons2021,Simons2023}). The CLEAR pointings overlap with the larger 3D-HST survey area \citep{Momcheva2016}, which provides slitless G141 grism spectra of 2-orbit depth with spectral coverage of 1.1-1.65~\micron.

These data will be described fully in the survey paper on the data release \citep{Simons2023} and have been discussed in other works using these data \citep[e.g.,][]{Estrada-Carpenter2019,Estrada-Carpenter2020,Simons2021,Jung2022,Backhaus2022a,Cleri2022,Matharu2022,Papovich2022,Backhaus2022b}.

\subsection{G102 and G141 Spectroscopy, Redshifts and Line Fluxes}

The \grizli\ (grism redshift and line analysis) pipeline\footnote{\texttt{https://github.com/gbrammer/grizli/}} serves as the primary method of data reduction for the CLEAR dataset. In contrast to traditional methods of extracting one-dimensional (1D) spectra from slit observations, \grizli\ directly fits the two-dimensional (2D) spectra with model spectra convolved to the galaxy image for multiple position angles of grism observations. This process yields complete and uniform characterization of the suite of spectral line features of all objects observed in each of the G102 and G141 grisms. The flux calibrations of the G102 and G141 spectra are, in general, accurate to within a few ($\sim$3) percent \citep{Estrada-Carpenter2019,Pirzkal2016,Pirzkal2017,Lee2014}. The most relevant of these spectral properties for our analysis are redshifts, line fluxes, and emission-line maps. The \nev\ doublet at 3346 and 3426 \AA\ is fit with a free ratio, i.e., \grizli\ does not force a ratio of $\nev\ \lambda 3426/\nev\ \lambda 3346 = 2.73$ (the expected ratio under typical nebular conditions, see Appendix \ref{app:pyneb}.)

In this work, we use the CLEAR v4.1 catalogs (\cite{Simons2021,Simons2023}). The data products of these catalogs include emission line fluxes, spectroscopic redshifts, and other derived quantities and their respective uncertainties for 6048 objects from \grizli\ run on the combination of the G102 and G141 grism data and broad-band photometry using the 3DHST+ catalogs. Of these galaxies, 4707 galaxies have coverage with both G102 and G141, which constitutes the initial catalog which we used to identify galaxies for our study here.  The emission-line fluxes from the \grizli\ reduction presented in this work are not corrected for attenuation by dust in the ISM.

The uncertainties of the emission lines account for the uncertainties of the continuum model as they are fit simultaneously. We note that the low spectral resolution may bias our sample to large equivalent widths, and the uncertainties in the continuum may lead to more uncertain equivalent widths than higher-resolution samples.

\subsection{Photometry and Derived Quantities}
We use stellar masses for objects in our sample from the 3D-HST catalog \citep{Skelton2014}, derived from the CANDELS photometry \citep{Grogin2011,Koekemoer2011}. The stellar masses are calculated by modeling the spectral energy distribution (SED) with FAST \citep{Kriek2009}, using a \cite{Bruzual2003} stellar population synthesis model library, a \cite{Chabrier2003} IMF, solar metallicity, and assuming exponentially declining star formation histories. The stellar masses of our galaxies are generally robust to these assumptions because the peak of the stellar emission is well-constrained by the high-quality CANDELS near-IR imaging. Stellar masses from the 3D-HST survey have a mass limit at $z \sim 2$ of $\log(M_*/M_\odot) \sim 8.5$ for $H < 25$ \citep{Skelton2014}.

To place these galaxies in context, we also compare their SFRs to other galaxies at similar redshifts.  For this purpose, we use UV continuum SFRs from the Cosmic Assembly Near-infrared Deep Extragalactic Legacy Survey (CANDELS)/ Survey for High-$z$ Absorption Red and Dead Sources (SHARDS) catalog of \cite{Barro2019}, which supplements the CANDELS multiwavelength data with SHARDS photometry \citep{Perez-Gonzalez2013} in GOODS-N and GOODS-S. Attenuation-corrected UV SFRs are calculated using the \cite{Kennicutt1998} calibration with a dust attenuation correction.  This is fully described in \cite{Barro2019}.
%

\subsection{Parent Dataset and Sample Selection} \label{sec:sample}
Our parent sample represents all CLEAR galaxies within the redshift range for detectable \nev\ and \oiii\ in the G102 and G141 spectrum ($1.39 < z < 2.40$). Requiring the wavelength coverage of a strong line such as \oiii\ eliminates many potentially spurious objects from the prospective sample and secures reliable spectroscopic redshifts for each object. The wavelength limits of this selection are set by the coverage of the blue end of the G102 and red end of the G141 grisms (8000 \AA\ and 16500 \AA, respectively). The CLEAR spectral extractions are limited to galaxies with $m_{\rm F105W} < 25$. 

We select a sample of \nev-emitting galaxies from the CLEAR parent catalog using the following steps:
\begin{itemize}
    \item Require a grism spectroscopic redshift, $1.39 < z < 2.30$, such that both \nev\ lines and \oiii\ are all within the observed-frame spectral range of G102 and G141 sensitivity (0.8-1.65~$\mu$m).
    \item Require signal-to-noise ratio of 3 for the stronger line of the \nev\ doublet (3426\AA) and \oiii.
    \item Visual inspection of direct images with 1D and 2D spectra.
\end{itemize}
This last step ensures that no objects with poor continuum-modeling and/or bad contamination subtraction make it into the final selection (this is a known issue with slitless spectroscopy, and visual inspection is important, especially for studies of objects with faint(er) emission lines like ours here, e.g., \citealt{Zeimann2014,Zeimann2015,Estrada-Carpenter2019,Estrada-Carpenter2020,Backhaus2022a,Backhaus2022b}). Each object was inspected by at least three authors. This selection rejects $\sim 40\%$ of objects which pass the first two criteria.

After applying all these selection processes, we have a sample of 25 \nev-emitting objects in CLEAR within the allowable redshift range of the G102 and G141 grisms ($1.39 < z < 2.30$). In our sample, all galaxies have \oiii\ \snr>5 (significantly greater than the minimum requirement of \snr>3). This lends greater credence to the spectroscopic redshifts and line identification. We also note that 9/25 objects have \nev\ \snr>5; the implications of the veracity of the \nev\ detections are discussed in Section \ref{sec:discussion}. Our sample comprises approximately 2.6\% of all objects in the CLEAR catalog in this redshift range. 

\begin{figure}[t]
\epsscale{1.1}
\plotone{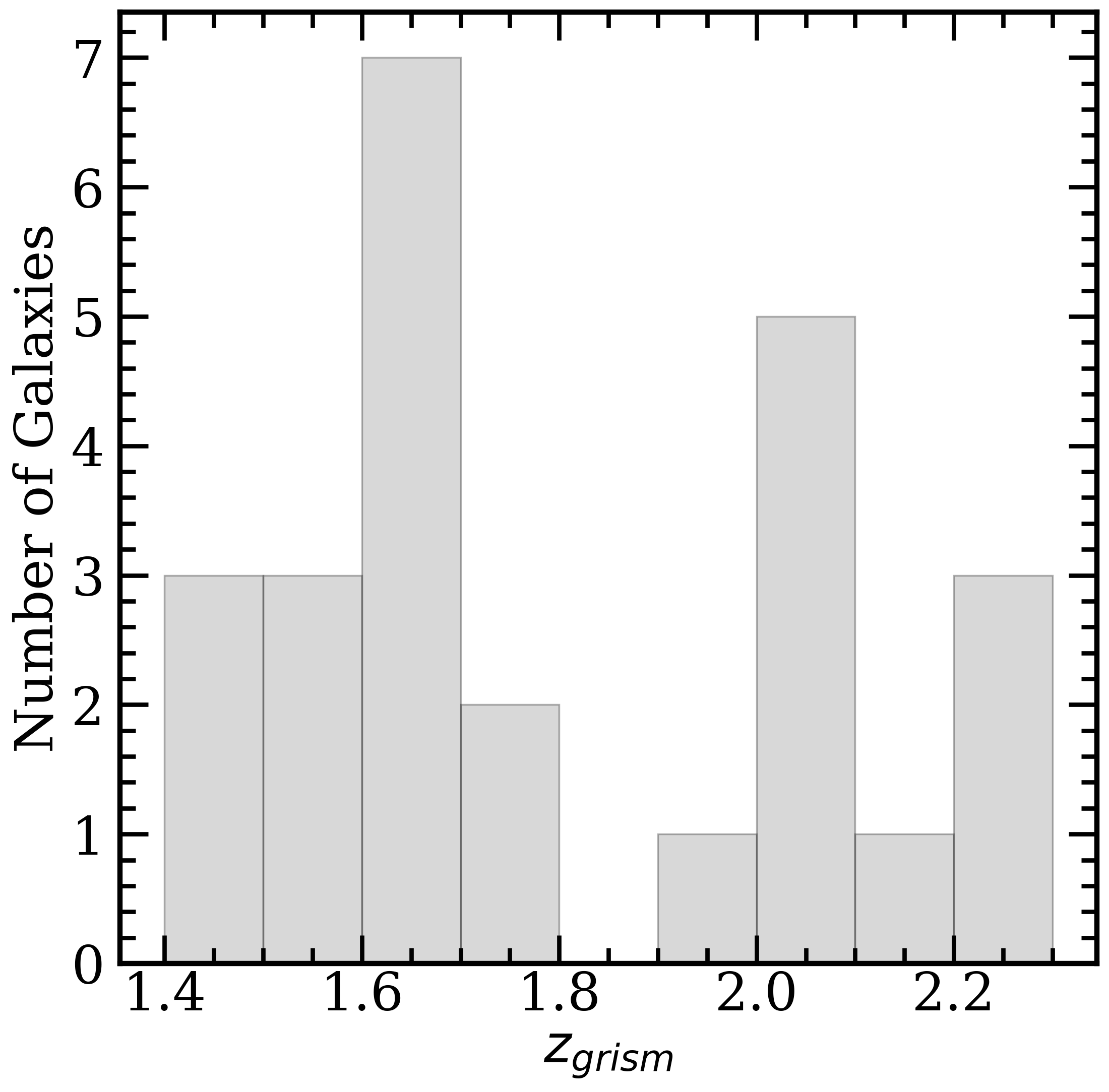}
\caption{Grism redshift distribution of our sample of 25 \nev\ galaxies. The G141 grism wavelength range limits the detection of \nev\ and \oiii\ to $1.39 < z < 2.30$. Our sample has a redshift range of $1.40 < z < 2.29$, with a median grism redshift of 1.61. The spike in sources at $z\sim 1.6$ is consistent with an overdensity of sources in GOODS-S at this redshift. 
\label{fig:zhist}}
\end{figure}

Our final sample has a redshift range of $1.40 < z < 2.29$, with a median grism redshift of 1.61. The redshift distribution of the 25 \nev-detected galaxies in our sample is shown in Figure \ref{fig:zhist}.  It is fairly evenly distributed in redshift, with a possible spike at $z\sim 1.6$, which corresponds to the known overdensity of galaxies at this redshift in the GOODS-S field \citep{Estrada-Carpenter2019}. The properties and derived quantities (including line fluxes) for the galaxies in our sample are shown in Tables \ref{tab:sample} and \ref{tab:lines}.

\begin{figure*}[t]
\epsscale{1.1}
\plotone{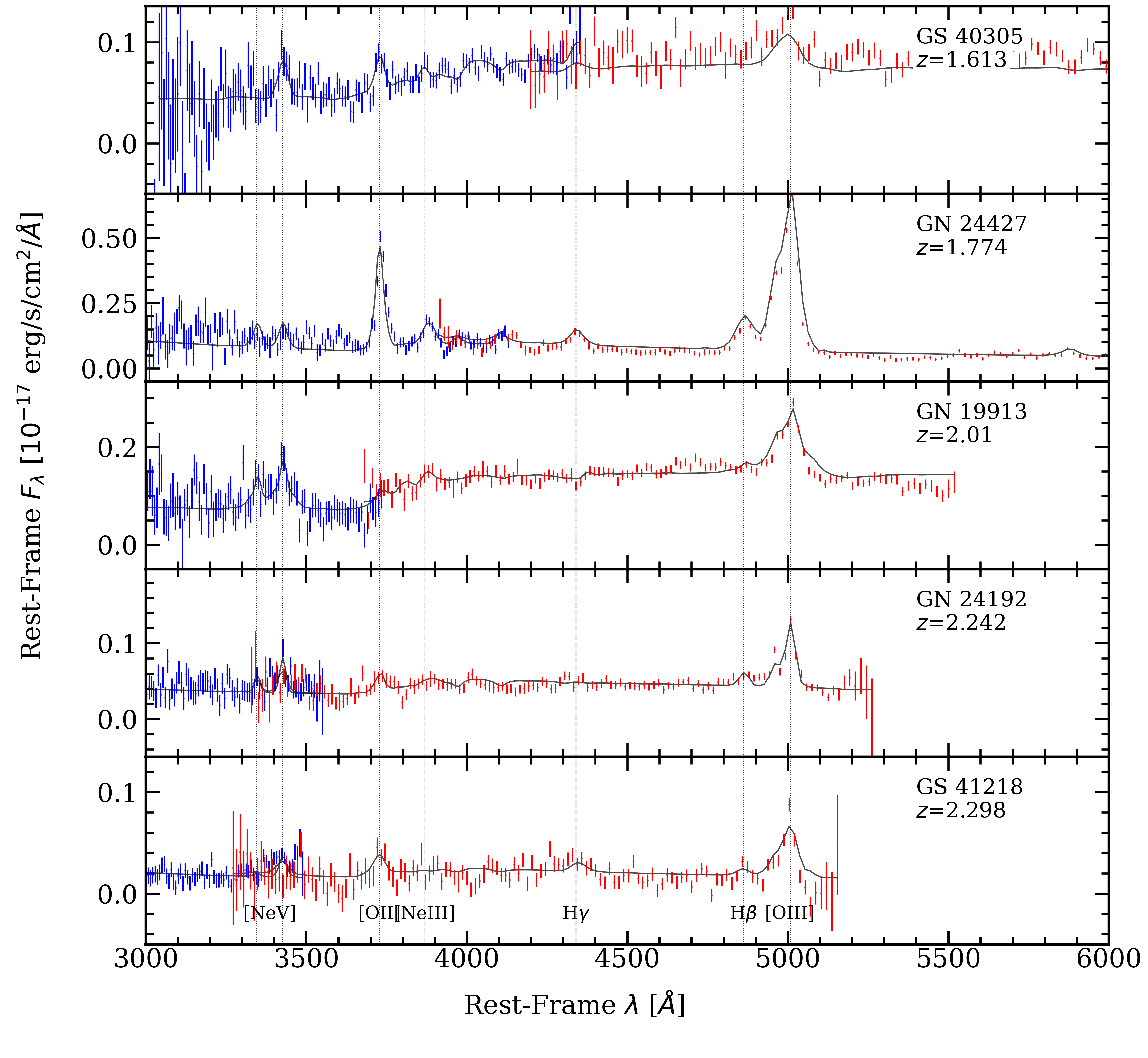}
\caption{Rest-frame one-dimensional spectra for five \nev-emitting objects in the CLEAR sample, ordered by increasing redshift. The G102 (blue) and G141 (red) spectra show the median points with 1$\sigma$ uncertainties from all exposures for this object. The dotted lines indicate emission features of interest: \nev\ $\lambda\lambda$3346,3426,\oii\ $\lambda\lambda$3726,3729, \neiii\ $\lambda$3869, H$\gamma$ $\lambda$4340, \hb\ $\lambda$4861,  and \oiii\ $\lambda\lambda$4959,5007.
\label{fig:spectra}}
\end{figure*}

Figure \ref{fig:spectra} shows rest-frame one-dimensional (1D) spectra of five \nev-emitting objects in the CLEAR sample, ordered by increasing redshift. The points and error bars shown in the 1D spectra are the medians in each bin of wavelength over all exposures. The points are separated into G102 (blue) and G141 (red). We note the region around several lines of interest: \nev\ $\lambda\lambda$3346,3426, \oii\ $\lambda\lambda$3726,3729, \neiii\ $\lambda$3869, H$\gamma$, \hb,  and \oiii\ $\lambda\lambda$4959,5007.

\begin{figure}[t]
\epsscale{1.1}
\plotone{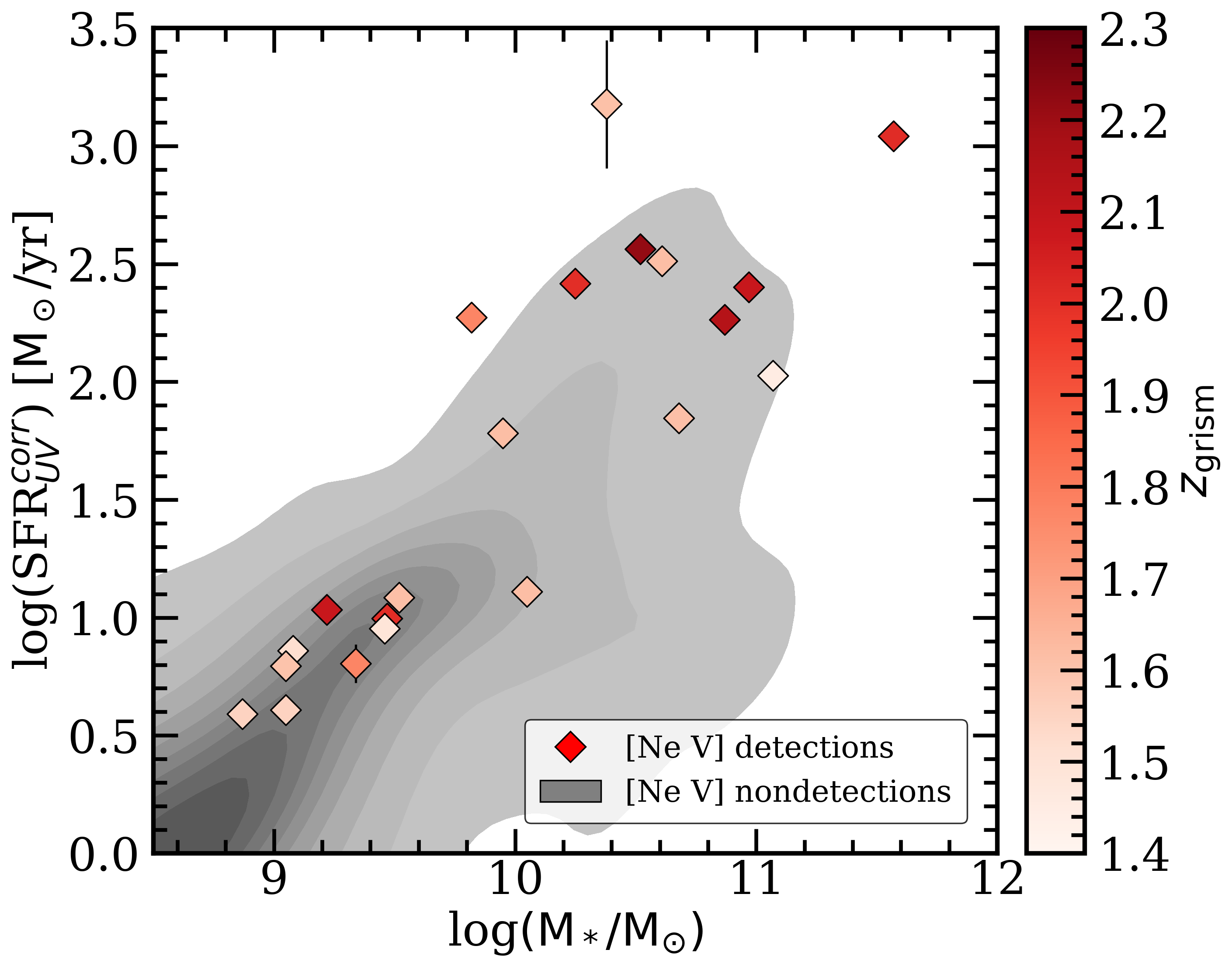}
\caption{The relation between attenuation-corrected UV star formation rate and stellar mass for galaxies of redshift $1.39 < z < 2.30$. Our sample of \nev-emitting galaxies is shown as diamonds color-coded by grism redshift, with the rest of the CLEAR galaxies with derived UV SFRs in this redshift range shown as gray contours. We show the 1$\sigma$ uncertainties in the attenuation-corrected UV SFRs for the \nev\ detections, most of which are smaller than the marker size. The lower-mass objects in our sample are broadly consistent with the rest of the CLEAR sample in SFR-mass space; however our sample includes several objects with higher mass / SFR than expected compared to the parent population, with 11/25 objects outside of the second to last (85\%) contour, 3 of which are outside of the last (95\%) contour.
\label{fig:sf_m}}
\end{figure} 


Figure \ref{fig:sf_m} shows the relation between star formation derived from attenuation-corrected UV luminosity and stellar mass, from \cite{Barro2019}. Our sample is again broadly consistent with CLEAR towards lower stellar mass. All but three of our \nev-emitting galaxies lie within the 95\% contours of the CLEAR parent population. Eight objects lie between the 85\% and 95\% contours, indicating that nearly half of the galaxies in our sample have elevated SFRs compared to the rest of CLEAR. One of these objects (GS 42758) has the highest attenuation-corrected UV SFR in the sample and has a V-band attenuation of 1.1 magnitudes \citep{Skelton2014}, and is also an X-ray AGN (see section \ref{subsec:xagn}). The rest of the sample is consistent with low-dust: we see a median V-band attenuation for the sample of 0.3 magnitudes.

\subsection{Comparison with X-ray AGN Catalogs}\label{subsec:xagn}
The CLEAR fields include the deepest X-ray imaging from \textit{Chandra} available anywhere on the sky (7 Ms in GOODS-S and 2 Ms in GOODS-N, see \citealt{Luo2017} and \citealt{Xue2016}, respectively). We use the X-ray luminosity for galaxies to diagnose their AGN activity. We matched our sample with the X-ray catalogs for the Chandra Deep Field-North and  -South (CDF-N and CDF-S) \citep{Xue2016,Luo2017}. Classifications from these catalogs include ``AGN'', ``galaxy'', or ``star'', where the AGN classification must satisfy at least one of the following four criteria from \cite{Xue2011}, that is, we combine these with a logical OR:
\begin{itemize}
    \item $L_{0.5-7~\mathrm{keV}}\geq 3\times10^{42}~\mathrm{erg~s}^{-1}$ (consistent with Luminous AGN)
    \item Effective photon index $\Gamma \leq 1.0$ (evidence for Obscured AGN)
    \item X-ray-to-optical flux ratio of $\log(f_X/f_R)>-1$ where $f_x = f_{0.5-7~\mathrm{keV}}, f_{0.5-2~\mathrm{keV}}$ or $f_{2-7~\mathrm{keV}}$ (evidence for AGN origin of X-ray emission)
    \item Excess X-ray emission over expectation from pure star formation $L_{0.5-7~\mathrm{keV}}\geq 3\times(8.9\times10^{17}~L_R)$
\end{itemize}
where $f_X$,$L_X$, $f_R$ and $L_R$ are the X-ray and $R$-band fluxes and luminosities, respectively. Objects labeled as ``galaxies'' in the X-ray catalogs are those which are confirmed to be galaxies (e.g., not stars/objects with a redshift of 0) but do not meet any of these criteria. This selection is subject to the caveat that luminous starbursts may be able to produce sufficient X-ray emission from X-ray binaries (XRBs) to be classified as AGN \citep{Lehmer2010}.

Matching with our \nev\ sample, we find 8 objects with X-ray detections classified as AGN. Throughout this work, we will denote these X-ray detected AGN in figures with a solid X marker where appropriate. This selection alone suggests that \nev\ is a useful tracer of AGN activity: 32\% (8/25) of the \nev\ sample are classified as X-ray AGN, while only 6.5\% of the rest of the CLEAR galaxies at this redshift range are classified as X-ray AGN.

\subsection{Comparison with IR AGN Catalogs}\label{subsec:iragn}
In addition to the X-ray matching to select potential AGN, we also select objects which are identified as AGN by mid-IR photometry. We use photometry from the Infrared Array Camera (IRAC) on \textit{Spitzer} \citep{Fazio2004,Lacy2004,Stern2005}. Our IR color selection criteria are outlined in \cite{Donley2012} and \cite{Coil2015}, designed to limit contamination by star-forming galaxies to $z<3$ while maintaining reliability and completeness. For the following selection, our notation is such that 
\begin{align}
    x = \log_{10}\left(\frac{f_{5.8\mu \mathrm{m}}}{f_{3.6\mu \mathrm{m}}}\right), \quad y = \log_{10}\left(\frac{f_{8.0\mu \mathrm{m}}}{f_{4.5\mu \mathrm{m}}}\right)
\end{align}
The selection of IR-AGN from \cite{Donley2012} requires all of the following criteria:
\begin{itemize}
    \item Objects are detected in all four IRAC channels (peak wavelengths $3.6\mu \mathrm{m}$, $4.6\mu \mathrm{m}$, $5.8\mu \mathrm{m}$, $8.0\mu \mathrm{m}$)
    \item $x \geq 0.08$ and $y \geq 0.15$ 
    \item $y \geq 1.21x - 0.27$
    \item $y \leq 1.21x + 0.27$
    \item $f_{8.0\mu \mathrm{m}} > f_{5.8\mu \mathrm{m}} > f_{4.5\mu \mathrm{m}} > f_{3.6\mu \mathrm{m}}$
\end{itemize}
This selection identifies 2 objects in our \nev-detected CLEAR sample as IR-AGN. Both objects identified through this photometric selection are also identified as X-ray AGN in the selection given in Section \ref{subsec:xagn}. IR-AGN selection generally samples more luminous AGN than X-ray selection \citep{Mendez2013}. Consequently, it is unsurprising to find that the two IR AGN in our sample are also X-ray detected. 

This selection, similarly to the X-ray AGN selection, suggests that \nev\ emission traces AGN activity: 10.5\% of the \nev\ sample are classified as IR AGN, while only 1.4\% of the rest of the CLEAR galaxies at this redshift range are classified as IR-AGN.



\section{Spectral Classification of Star Formation and AGN Activity}\label{sec:sf_agn}

To characterize the source of ionizing radiation for each galaxy in our \nev\ sample, we primarily use the \oiii/\Hb\ ratio combined with other diagnostics. When \oiii/\Hb\ ratios are compared to other galactic parameters (e.g. stellar mass and other line ratios), the relation can be used to diagnose the ``activity'' of the galaxy: ``active'' galaxies hosting AGN or ``inactive'' galaxies dominated by star formation.

\subsection{The Mass-Excitation Diagram}
We start by considering the ``mass-excitation'' (MEx) diagram, which combines the \oiii/\Hb\ ratio with the stellar mass \citep{Juneau2011,Juneau2014}. The MEx diagram is the most inclusive of these AGN diagnostics, i.e., it is suitable for galaxies even in the case that we have only one line ratio, \oiii/\Hb.  The dependence of the MEx diagnostic on stellar mass includes biases and assumptions of SED modeling \citep{Barro2019}. \cite{Juneau2014} derived empirical demarcations between AGN and galaxies with star-formation.  First, they set a relation between \oiii/\Hb\ and  and stellar mass that identifies galaxies with ionization from only AGN, defined to have $y\equiv \log(\oiii/\Hb)$\footnote{For the \hst\ G102 and G141 grism spectral resolution, the $\oiii~\lambda\lambda 4959,5007$ lines are blended, so we use the blended flux for these analyses.} greater than the value given by the relationship
\begin{equation}\label{eq:mex_top}
    y = \left\{
        \begin{array}{ll}
            \frac{0.375}{(x - 10.5)} + 1.14, & \quad x \leq 9.9 \\
            \\
            410.24 - 109.333x \\
            + 9.71731x^2 - 0.288244x^3, & \quad \text{otherwise}
        \end{array}
    \right. 
\end{equation}
where $x\equiv\log(M_*)$. Second, they use a relation between line ratio and stellar mass, where galaxies below this line ratio contain ionization from only star-formation, where $y$ for this relation is given by 
\begin{equation}\label{eq:mex_bottom}
    y =  352.066 - 93.8249x + 8.32651x^2 - 0.246416x^3
\end{equation}
in the range $9.9<\log(M_*)<11.2$. To summarize, galaxies which lie above the top curve (Eqn.~\ref{eq:mex_top}) in the MEx diagram are classified as AGN; galaxies which lie below the bottom curve (Eqn.~\ref{eq:mex_bottom}) are classified as star-forming. We further define galaxies that lie between these two curves as composite sources, with contributions from both AGN and star-formation.

However, we need to adjust the MEx diagram to account for redshift evolution.  For example, \cite{Coil2015} studied a population of AGN at $z\sim 2.3$ with rest-frame optical emission line ratios.  They concluded that they needed to apply a shift of $\Delta\log(\mathrm{M_*/M_\odot}) = 0.75$ dex, which more accurately separates star-forming galaxies and AGN at this redshift.  This provides a more pure selection of confirmed X-ray AGN via the MEx diagram than the local \cite{Juneau2014} line.

Here, we use the MEx relation and adopt a shift intermediate between that of \cite{Juneau2014} and \cite{Coil2015} as the CLEAR \nev\ sample lies at a median redshift $z\sim 1.61$, between that of the samples of these other studies.  We construct a simple empirical model to encapsulate this shift in the MEx relation from redshift evolution.  This expands on the 0.75 dex shift from \cite{Juneau2014} to \cite{Coil2015}, which becomes: 
\begin{equation}\label{eq:mex_z} 
x = \log(M_*) + 0.2(1+z)
\end{equation}
 where $x$ is defined above, and $0.2(1+z)$ represents the shift in the x-axis from the \cite{Juneau2014} line (Equations \ref{eq:mex_top} and \ref{eq:mex_bottom}). We arrive at this shift of $0.2(1+z)$ as it classifies all X-ray-confirmed AGN with stellar mass $\log(M_\star/M_\odot) > 9$ as AGN or 1$\sigma$ consistent with the AGN/SF dividing line. This shift keeps the same purity of the \cite{Coil2015} AGN selection, but (as we discuss below) this is more consistent with the X-ray detected \nev\ sources in our sample which would otherwise be labeled star-forming by the \cite{Coil2015} line.

Figure \ref{fig:agn_sf} (left panel) shows the MEx diagnostic for the galaxies in our CLEAR \nev\ sample (where we include all of our galaxies that have S/N $>$ 1 in \Hb\ and \oiii). The diamonds show the 18 sources in our \nev\ sample that satisfy this requirement. We denote galaxies in our CLEAR \nev\ sample detected in X-rays with a ``thin X'' marker, and we denote galaxies that satisfy the IR-AGN definition with a ``hollow X'' marker. We also show on the MEx diagram those galaxies detected with S/N $>$ 1 in CLEAR without \nev\ detections, in the same redshift range as the \nev\ sample, as small gray points.

Roughly half of the sources in our \nev\ sample lie in the AGN region of the MEx plot. If we compare to the MEx definition defined by \cite{Juneau2014} (J14)  then 12 / 18 (=66\%) of the CLEAR \nev\ sample show evidence of AGN ionization.  This includes all of the sources detected in X-rays or identified as IR-AGN. Using the MEx definition of \cite{Coil2015} (C15), this fraction drops to 8/18 (44\%), and misses two of the X-ray sources, but includes all the IR-AGN. When we compare the MEx selection using the redshift offset for our \nev\ sample (assuming the median redshift of the sample $z=1.61$ with Eqn.~\ref{eq:mex_z}, as indicated by the solid lines in the MEx panel of Figure~\ref{fig:agn_sf}) we would select 11/18 (61\%) of the \nev\ sources including all the X-ray sources and IR-AGN. 

The results of this MEx analysis are broadly consistent with previous work done with \nev\ galaxies. \citealt{Mignoli2013} finds an AGN fraction of $\sim$80\% of \nev\ emitting galaxies at $z\sim0.8$ are consistent with AGN classification via MEx, although the analyses are not directly comparable, as the \citealt{Mignoli2013} uses the \citealt{Juneau2011} $z\lesssim0.1$ MEx division and does not perform the same redshift evolution of the MEx line as in this work. Given the relatively small sample in this work, as well as the biases of MEx and stellar mass derivations, we caution use of this diagnostic when others are available (see Section \ref{sec:discussion} for more discussion).

\subsection{The ``OHNO'' Diagram}
We next explore other emission line diagnostics designed to separate galaxies with ionization from AGN and star-formation.  One such diagnostic is the \oiii/\Hb\ and \neiii/\oii\ (the ``OHNO'') diagram \citep{Zeimann2015,Backhaus2022a}. This diagnostic compares ratios of emission lines at similar wavelengths (\oii\ $\lambda\lambda$3726,3729 \AA, \neiii\ $\lambda$3869 \AA, \hb\ $\lambda$4861 \AA,  and \oiii\ $\lambda\lambda$4959,5007) where the production of \oiii\ and \neiii\ both require higher photon energies: the ionization energy of O$^0$ is 13.6 eV, that of O$^+$ is 35.1 eV, and that of Ne$^{+}$ is 41.0 eV. Galaxies with strong \oiii/\Hb\ and/or \neiii/\oii\ require harder radiation fields, typically found in the emission-line regions of AGN.  \cite{Backhaus2022a} showed that division in the OHNO line ratios separates X-ray--selected AGN from non-AGN (based classifications from the deep X-ray data in the CDF-N and CDF-S fields).

\begin{figure*}[t]
\epsscale{1.15}
\plotone{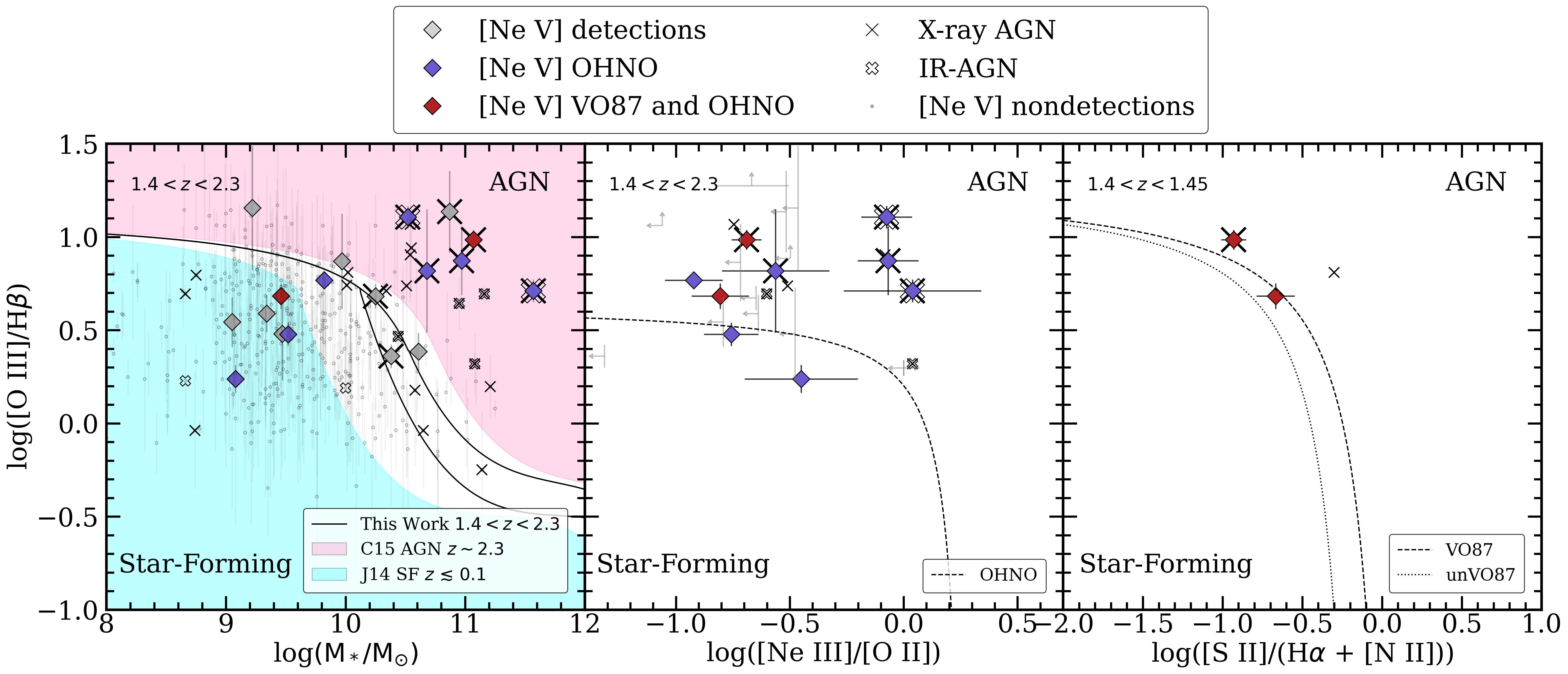}
\caption{Emission line diagnostics of star-formation/AGN activity. Each panel requires \snr>1 for all represented emission lines. Galaxies which lie above the respective dividing line are classified as AGN, and galaxies which lie below the line are classified as star-forming. X-ray AGN are shown with a solid X, and IR-AGN are shows with a hollow X. \nev-detected objects are colored by their detected emission line pairs: gray diamonds have detections in \oiii/\hb\ only, dark purple diamonds have \oiii/\hb\ and \neiii/\oii\ (OHNO), and red diamonds have \oiii/\hb, \sii/[\Ha\ + \nii] (VO87).  Redshift ranges for the coverage of all respective lines are shown in the top left of each panel. \textit{Left}: Mass-excitation diagram: the relation of log(\oiii/\Hb) vs. stellar mass. Galaxies in CLEAR that are undetected in \nev\ in this redshift range  (\nev\ nondetections) are shown as small grey points. The blue and pink shaded regions show the local \cite{Juneau2014} star-forming and $z\sim2.3$ \cite{Coil2015} AGN regions, respectively.  The $z\sim1.6$ redshift-evolved MEx line from Equation \ref{eq:mex_z} is shown in black. \textit{Center}: the OHNO diagram using log(\oiii/\Hb) versus log(\neiii/\oii). We also show the limits of line ratios for objects in the cases where objects are undetected in various permutations of \Hb, \neiii, or \oii. \textit{Right}: The VO87 diagram for log(\oiii/\Hb) versus log(\sii/[\Ha\ + \nii]). The dashed line shows the \cite{Veilleux1987} line for $z\sim 0$, and the dotted line shows the \cite{Backhaus2022a} ``unVO87'' dividing line for galaxies at $z\sim 1$. The limited redshift range for the detection of all five of these lines leaves much smaller samples than other diagnostics. Both of the galaxies with VO87 lines also have detected \neiii/\oii. Both the points with well detected line ratios and limit behaviors of other \nev\ detections suggest a broad preference for \nev\ detections to be classified as AGN in all three of these diagnostics. Based on these line diagnostic plots, the CLEAR \nev-emitting galaxies are broadly consistent with ionization from AGN.
\label{fig:agn_sf}}
\end{figure*} 

The center panel of Figure \ref{fig:agn_sf} shows the OHNO diagram for the galaxies in our CLEAR sample. The redshift range which allows for all five of these lines in the \hst\ G102 and G141 grism coverage is \zohno. Nine of our \nev-detected objects are well-detected in the four OHNO lines (using the OHNO AGN/star-formation separation from \citealt{Backhaus2022a}).  Of these, 8 (out of 9 = 89\%) of the galaxies have line ratios consistent with ionization of AGN. This includes all five (100\%) of the X-ray--detected \nev\ sources in our sample. There is one galaxy in our \nev-emitter--sample that falls below the AGN line in the OHNO diagram, but it is consistent with being an AGN within its $1\sigma$ uncertainties based on the classification line defined by \cite{Backhaus2022a}.  Therefore, for the \nev-emitting galaxies in our sample that we can place on the OHNO diagram, all but 1 galaxy are 1$\sigma$ consistent with ionization from AGN. The single object which lies greater than 1$\sigma$ outside of the AGN region of the OHNO diagram has large horizontal error bars due to a low \snr\ \oii\ detection (we therefore cannot rule out ionization from an AGN in this object).

In the OHNO panel, we also show the limiting cases for objects in our sample which do not have well-detected \Hb, \neiii, or \oii\ emission lines, in various different permutations of undetected lines (i.e., those not detected in \Hb, \neiii, \oii, or some combination thereof). This analysis shows that even the \nev-detected objects without all of the necessary lines preferentially lie in the AGN region.

%
\subsection{The ``unVO87'' Diagram}
Another diagnostic used to separate AGN and star-forming galaxies is the relation between \oiii/\Hb\ and \sii\ combined with \Ha\ \citep[][VO87 hereafter]{Veilleux1987}.    The VO87 diagram has been applied to many studies of galaxies (including AGN and star-formation) at $z\sim 0$ \citep{Veilleux1987,Kauffmann2003,Kewley2001,Kewley2019a,Trump2015}. The original VO87 relation to divide AGN and star formation is given by 
\begin{equation}\label{eq:VO}
\log\left(\frac{\oiii}{\Hb}\right)=\frac{0.48}{\log( \sii/\Ha) + 0.10}+1.3
\end{equation}
At the resolution of the \hst/WFC3 grisms, the \sii\ lines are blended with each other, and \Ha\ is blended with \nii.  We therefore use the ``unresolved'' VO87 relation (henceforth ``unVO87'') which has been tested at $z\sim 1$ for galaxies where these lines are blended (unresolved, see \citealt{Backhaus2022a}). In this case, \cite{Backhaus2022a} define an empirically derived relation to separate AGN and star-forming galaxies as follows: 
\begin{equation}\label{eq:unVO}
\log\left(\frac{\oiii}{\Hb}\right)=\frac{0.48}{\log\left(\sii/[\Ha+\nii]\right)+0.12}+1.3
\end{equation}
where galaxies lying above the curve (higher \oiii/\Hb) are classified as AGN and those below the curve are classified as star forming for both the \cite{Veilleux1987} and \cite{Backhaus2022a} curves.

Figure \ref{fig:agn_sf} (right panel) shows the unVO87 diagram for objects in CLEAR in the (rather) narrow redshift range which allows for \hst\ G102 and G141 grism coverage of \nev\ along with all of \oiii, \Hb, \sii, and \Ha\ (\zvo). We show both the original $z\sim 0$ VO87 relation and the $z\sim 1$ unresolved unVO87 AGN/SF dividing lines. Given the very limited redshift range to allow for all four lines needed for unVO87 and \nev, there are only two \nev-emitting objects in this subsample.  These are shown in red. One of the \nev-detections in this panel is categorized as an X-ray AGN. The X-ray undetected \nev-emitter is consistent with the unVO87 division within its uncertainties. There is also one object in the CLEAR sample that is undetected in \nev\ in this redshift range which is identified as an X-ray AGN, and is classified as such by both the VO87 and unVO87 divisions.


As these emission line ratio diagnostics suggest, the \nev\ sources in CLEAR appear primarily consistent with ionization from AGN. This is clear (pun intended) from the MEx diagram, and OHNO and VO87 relations in Figure~\ref{fig:agn_sf}, where the majority of the \nev\ sources fall in regions consistent with ionization from AGN.  There are three objects which have contradictory classifications between the MEx and OHNO/VO87 diagnostics, but we favor the classifications from the emission-line diagnostics (OHNO and VO87) as they are not subject to the biases and uncertainties involved with the estimation of stellar masses from SED fitting. In total, all but 4 (21/25 = 87.5\%) of the \nev-detected objects in our sample are consistent with AGN classification from these three diagnostics, either by their detected line ratios or limiting behaviors. We discuss the implications of the results of these analyses in Section \ref{sec:discussion}.
%
%

\begin{figure*}[t]
\epsscale{1.1}
\plotone{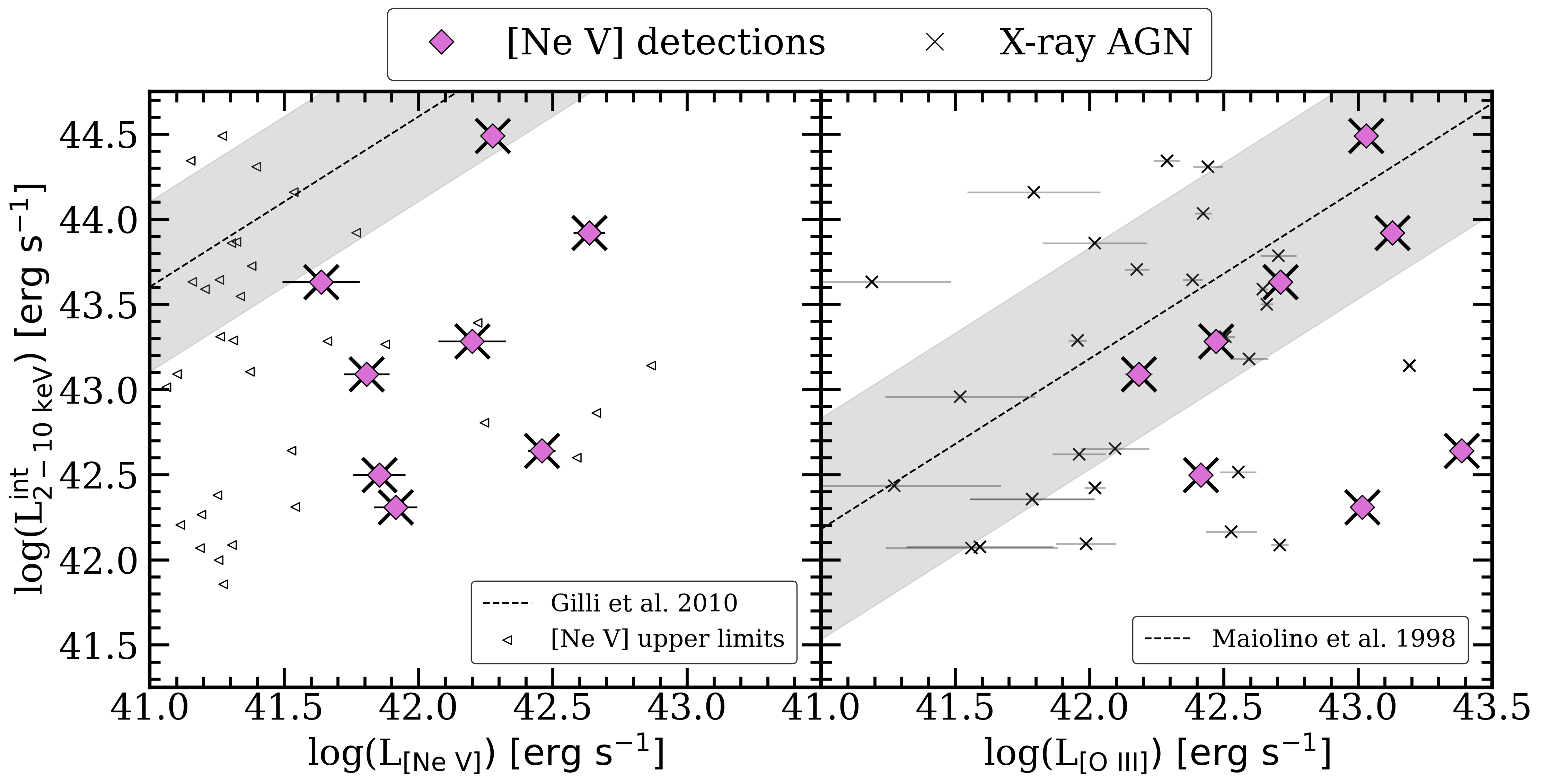}
\caption{The relation between intrinsic X-ray luminosity and \nev\ luminosity (\textit{left}) and \oiii\ luminosity (\textit{right}) for objects in CLEAR matching to the \cite{Xue2016} and \cite{Luo2017} X-ray catalogs, in the redshift range of our \nev\ sample. \nev-detected objects are shown as purple diamonds. We also show the 1$\sigma$ upper limits for the \nev\ nondetections as left-facing black triangles. The black dashed lines and gray shaded regions in each panel show the median and 1$\sigma$ relations for local Seyferts \citep{Gilli2010,Maiolino1998}. Our emission-line selected sample has preferentially higher \nev\ compared to X-ray luminosities than local Seyfert relations suggest. 
\label{fig:xnevoiii}}
\end{figure*} 


\section{Using X--ray Emission to Characterize \nev-Emitting AGN} \label{sec:nev_x}
In this section, we investigate the properties of \nev-emitting galaxies in relation to their X-ray emission. We explore the use of observed $\nev \lambda 3426$ \AA\ luminosities to probe AGN activity missed by other AGN selection methods, such as X-ray, IR, and emission-line diagnostics.

For the following analysis, we use luminosities of both X-ray detections and upper limits in X-rays for galaxies that are undetected (``X-ray nondetections''). We take X-ray fluxes from the Chandra Deep Field North (CDF-N) and South (CDF-S) catalogs (\cite{Xue2016} and \cite{Luo2017}, respectively). We calculate the upper limits of the luminosity of the X-ray nondetections using the hard-band flux detection limits from \cite{Xue2016} and \cite{Luo2017} of $5.9\times 10^{-17}\mathrm{erg~s}^{-1}\mathrm{cm}^{-2}$ and $2.7\times 10^{-17}\mathrm{erg~s}^{-1}\mathrm{cm}^{-2}$, respectively, and the grism redshifts from the CLEAR catalog. We perform a K-correction assuming a typical effective photon index of 1.8 \citep{Liu2017,Yang2016}. We transform our 0.5-7 keV luminosities to 2-10 keV luminosities for comparisons with other samples following \cite{Yang2016}, where $L_{2-10~ \mathrm{keV}} = 0.721 L_{0.5-7~ \mathrm{keV}}$.

In Figure \ref{fig:xnevoiii} we present the intrinsic X-ray luminosities for the 8 \nev-emitting galaxies classified as X-ray AGN by the \cite{Xue2016} and \cite{Luo2017} CDF catalogs, in relation to the \nev\ and \oiii\ luminosities. In the left panel, we also show the 1$\sigma$ upper limits of the \nev\ luminosity for X-ray AGN in the CDF catalogs not detected in \nev. We show the local Seyfert X-ray vs \nev\ and \oiii\ luminosity relations from \cite{Gilli2010} and \cite{Maiolino1998}, respectively. 

Our first result is that the majority of the CLEAR \nev-emitter galaxies have \nev\ luminosities that exceed the local scaling relation. Figure~\ref{fig:xnevoiii} shows that only two of the eight \nev-detected X-ray AGN are consistent (within 1$\sigma$) with the \nev--X-ray relation of \cite{Gilli2010} (which was derived from the observed 2-10 keV luminosities from local Seyferts). However, five of the eight are consistent with local \oiii--X-ray relation (derived from local Seyferts, \citealt{Maiolino1998}).   We do not find anything that differentiates the three galaxies that are outliers on both the \nev-/X-ray and \oiii-/X-ray relations from the rest of the sample:  these three galaxies show no special features in their properties nor spectra. We therefore conclude that AGN span a larger variation in \nev\ and X-ray emission than suggested from local Seyfert samples.  

Our sample is biased to high \nev\ luminosities by selection, which may be in part responsible for this result. However, we note that there are several objects which are consistent with the local X-ray to \nev\ and \oiii\ relations. 

If this result was due to an insufficient absorption correction to the observed X-ray luminosities, we would expect all of the objects from our sample to lie below the local relations. Instead, the large scatter in the X-ray luminosities of objects in our sample indicates that this is not the case: an additional flat correction needed to bring the low X-ray luminosity objects to the local relations would skew the objects with higher X-ray luminosities above the local relations. While a flat correction to all luminosities is likely not a perfect prescription, the objects which are discrepant from the local relations are uniformly distributed across all luminosities.

Previous studies have used the X-ray/\nev\ luminosity ratio to study AGN.  \cite{Gilli2010}  argued that the X-ray/\nev\ luminosity ratio ($L_X/L_{\nev}$) could be a useful indicator of CT AGN.  They observed that  all Seyferts in their sample with $L_X/L_{\nev} <~15$ showed evidence of CT AGN.  However, \cite{Gilli2010} assumed that AGN have a near constant intrinsic $L_X/L_{\nev}$ ratio, in which case the lower observed $L_X/L_{\nev}$ ratios imply obscuration of the X-ray emission. 

\begin{figure}[t]
\epsscale{1.1}
\plotone{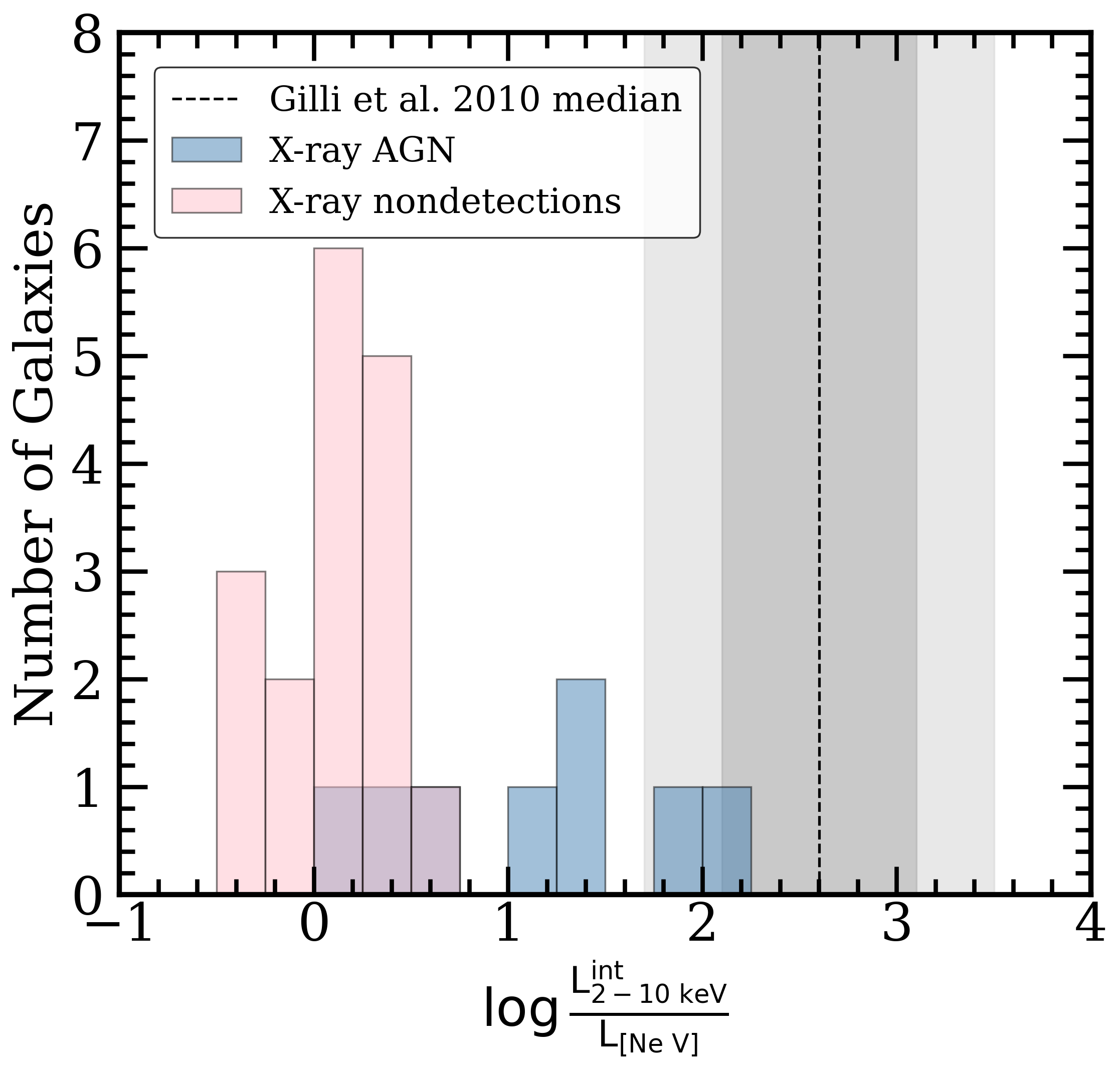}
\caption{The distribution of the intrinsic X-ray/\nev\ luminosity ratio for the 8 X-ray detected (blue) and upper limits for the 21 X-ray nondetected (pink) in our \nev-detected sample. The dashed black line and gray shaded regions show the median, 1$\sigma$, and 90\% ranges of the unobscured Seyferts in the \cite{Gilli2010} local sample. All but two objects in our sample lie below the 90\% lower limits of the unobscured local Seyferts.
\label{fig:xnev}}
\end{figure}

Figure \ref{fig:xnev} shows the distribution of the X-ray/\nev\ luminosity ratio for our sample, with the X-ray detections in blue and upper limits for the non-detections in pink. We also show the median ratio for the local ($z < 0.1$) Seyferts  from \cite{Gilli2010}.  The gray-shaded regions show the inter-68 and inter-90 percentiles for the \cite{Gilli2010} sample. 

The X/\nev\ distribution for the CLEAR \nev-emitter sample is systematically lower than that of the low-redshift Seyferts from \cite{Gilli2010}.  Only two of the objects in our CLEAR \nev-emitter sample have $L_X$/\nev\ ratios consistent (within the 90 percentile) with those of \cite{Gilli2010}.   The majority of the galaxies in our CLEAR \nev-emitter sample --- including four of the X-ray-detected galaxies, and all 21 of the X-ray \textit{non-}detected galaxies --- have X-ray/\nev\ ratios below the canonical value of $L_X$/\nev~$< 15$ used to identify CT AGN \cite{Gilli2010}. However, none of the \nev-emitter galaxies in our sample are consistent with being CT AGN given their absorption column densities (i.e., column densities $\log N_H < 24$, \cite{Li2019}). Our sample has a range of column densities $21.45 < \log N_H < 23.95$. We therefore conclude that X-ray/\nev\ does not uniquely identify CT AGN.

\begin{figure}[t]
\epsscale{1.1}
\plotone{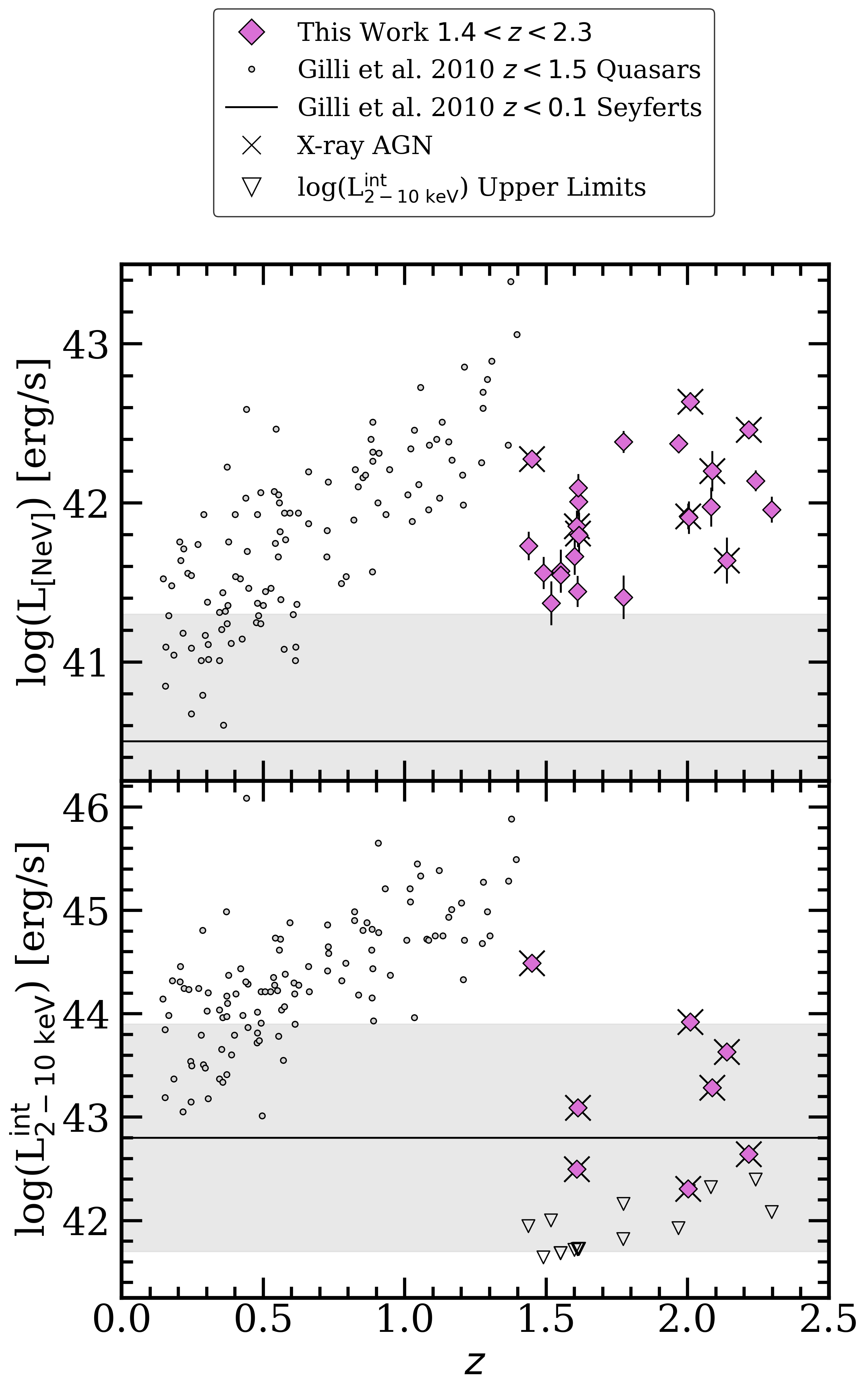}
\caption{The relations of \nev\ (\textit{top}) and intrinsic X-ray (\textit{bottom}) luminosity functions for the 8 X-ray AGN in our CLEAR sample (purple diamonds) and 112 lower redshift QSOs ($z\lesssim 1.5$) from \cite{Gilli2010} (gray circles). We also show the median of the X-ray and \nev\ luminosities of the \cite{Gilli2010} local ($z<0.1$) Seyferts sample as a black line, with the gray shaded region showing the $1\sigma$ scatter. We show 1$\sigma$ uncertainties which may be smaller than the size of the markers. Our higher redshift \nev\ selection probes a parameter space not observed in the X-ray selected samples, with X-ray luminosities comparable to local Seyferts and higher \nev\ luminosities more typical of $z\sim1$ QSOs. 
\label{fig:x_nev_z}}
\end{figure}

One reason the X-ray/\nev\ ratio is unable to identify CT AGN may be because there are systematic differences in our CLEAR \nev-emitter sample and other X-ray-selected samples of AGN. In particular, our sample of high-redshift galaxies spans different luminosities and redshift. Figure \ref{fig:x_nev_z} shows the distribution of \nev\ luminosities and intrinsic X-ray luminosities as a function of redshift for the eight galaxies in our \nev-emitter sample detected in X-rays.  The Figure compares these to samples of lower redshift ($z < 1.5$) luminous QSOs and very low-redshift ($z< 0.1$) Seyferts \citep{Gilli2010}.  For completeness, we include in the Figure those galaxies in our CLEAR \nev\ sample that are \textit{un-}detected in X-rays using the same prescription as those in Figure \ref{fig:xnev}.

Figure~\ref{fig:x_nev_z} shows that the CLEAR \nev--emitter sample has \nev\ luminosities consistent with QSOs from $0 \lesssim z \lesssim 1$, but has X-ray luminosities that are much lower, and more consistent with the range of X-ray luminosities seen in local Seyferts. This is evidence that there exists greater variation between the X-ray engine in AGN (presumably the accretion disk) and the narrow line region (which is responsible for the \nev\ emission).  We discuss this variation and its implications of the results of these analyses in Section \ref{sec:discussion}. 

\section{Discussion}\label{sec:discussion}

\subsection{The Nature of \nev\ Galaxies through Emission-Line Ratio Diagnostics}
The emission line ratio diagnostics in Figure \ref{fig:agn_sf} have several implications for the nature of \nev-emitting galaxies. In each of the three diagnostics (MEx, OHNO, and VO87), the \nev-detected objects are consistent with ionization from AGN: the MEx diagnostic classifies 12/18 (67\%) of \nev-detections as AGN, while the more reliable, yet less inclusive, OHNO diagram classifies 8/9 (89\%) of \nev-detections as AGN.

The presence of an AGN is expected given that the energy needed to produce \nev\ is 97.11 eV.  The ionizing spectra of stellar populations are not expected to produce copious radiation at this energy \citep{Olivier2022}.  The presence of an AGN is supported by the fact that a large fraction ($\sim$32\%) of the \nev\ galaxies are detected in X-rays and/or selected as AGN from their IR. This is especially true for the \nev\ sources in the high-excitation (AGN) regions of the plots in Figure~\ref{fig:agn_sf}. Therefore, given the coincidence of AGN among the \nev-selected galaxies, AGN appear to explain the origin of the \nev\ emission in most of our sample here. 

It is noteworthy, however, that there appears to be a population of \nev-detected galaxies in CLEAR that have \oiii/\Hb\ ratios below the threshold of traditional AGN selection.  This includes six sources in the MEx diagram of Figure~\ref{fig:agn_sf} with lower \oiii/\Hb\ ($\lesssim 1$), lower stellar masses ($\log M_\ast/M_\odot < 10$), and are undetected in the X-rays. While these objects have formally-detected \nev\ (>3$\sigma$), it is important to note that these six objects are among the weaker \nev\ detections in our sample (3<\snr<5). To account for the \nev\ emission in these galaxies requires some mechanism other than a bright AGN. This could include one or more of the following: 
\begin{enumerate}
    \item  Heavily obscured AGN:  It is possible these \nev-emitters contain deeply obscured AGN such that the X-ray emission is undetectable (even in the 2 or 7 Ms-depth data available for the CDF-N and -S fields).  Furthermore, obscured AGN should be detected as IR AGN \citep{Donley2012}, where again, the CDF-N and -S fields have some of the deepest far-IR imaging on the sky (\citealt{Barro2019,Guo2013} and references therein).  The lack of indications of AGN activity in the X-rays nor IR-emission in these galaxies disfavors this interpretation. 
    \item Weak AGN:  The \nev\ emission could stem from weak AGN, where again such objects would have X-ray emission below the detection limit for our sample.  This is an intriguing possibility, especially given the lower stellar masses for these galaxies. It is possible they host (lower-mass) intermediate mass black holes (IMBHs) with lower accretion rates, but are still able to produce a strong narrow-line region \citep{Greene2020}. Additional study of these galaxies for other high-ionization emission lines will be able to confirm this possibility (e.g., with rest-frame optical/near-IR spectroscopy from \jwst). A related potential explanation for the low observed \oiii/\hb\ ratios in these objects is through star formation coincident with the AGN phase boosting the \hb\ fluxes, thus hiding the AGN activity given these typical diagnostics.   
    \item Shocks or other extreme mechanisms:   \nev\ emission has been detected in several low-mass, nearby galaxies, where those studies argue the emission is produced by energetic shocks from supernovae or extreme stellar populations in a lower metallicity, high-density ISM \citep{Thuan2005,Izotov2012,Izotov2021,Olivier2022}. \citealt{Leung2021} also finds that \nev\ may be produced in higher-metallicity objects that have (in at least one case) indications of shocked gas from AGN-driven winds. It is plausible that some of the \nev\ emission in the lower-mass galaxies in our sample stem from similarly-produced shocks.  For this to be the case, we would expect to see indications of high density, which could be traced by resolved \sii\ or other density-sensitive lines (see Appendix \ref{app:pyneb}). Currently the \hst/WFC3 grism data has insufficient resolution to study these lines at these redshifts, but this would be possible with future \jwst\ spectroscopy at higher spectral resolution. 
\end{enumerate}

\subsection{The Nature of \nev\ Galaxies via their X-ray Emission}
The study of \nev\ in conjunction with observed-frame 2-10 keV luminosities offers insight into the relative amount of emission in the ``hard UV''/``soft X-ray'' regime (energies around $\sim$100 eV) in the spectra of these galaxies. The \nev\ emission therefore provides information unavailable from studies of AGN using only X-ray or IR.     As noted above (in Figure~\ref{fig:x_nev_z}), our sample of \nev-emitter galaxies in CLEAR have \nev\ emission similar to $z\sim 1$ QSOs but X-ray luminosities similar to local Seyferts.   This means that the galaxies in our \nev-emitter sample have a lower X-ray/\nev\ luminosity ratio than seen in other samples.   We discuss here our interpretation of the conditions for the lower X-ray/\nev\ ratios.  Specifically, this must be a result of either (1) reduced X-ray emission and/or (2) enhanced \nev\ emission in these higher redshift objects. 


The preferentially low X-ray/\nev\ ratios of our sample suggest an excess of $\sim$0.1 keV photons compared to the emission at $>1$ keV emission.  Enhanced \nev\ emission could be caused by several effects (or a combination of effects).  Strictly speaking, it requires a higher density of $\sim100$ eV photons. This could result from different geometry of the NLR and accretion disk \citep{Trump2011}, or conditions that conspire to enhance the emission of these ``soft X-ray'' photons compared to the ionizing spectrum of local objects. This could also be a result of excess \nev\ from shocked gas from AGN winds \citep{Leung2021}.
%
%

If geometry is the culprit of the enhanced \nev\ emission, then specific conditions seem to be required. For example, the enhanced \nev\ could be explained by anisotropy in the X-ray emission \citep{Yang2020} such that the NLR is illuminated by the accretion disk, but the sightline to the central engine is obscured.  However, this obscuration would likely be seen in the X-ray absorbing column, where the absorption column densities, $N_H$, of our sample suggest that our objects are \textit{not} CT (i.e. log~$N_H < 24$) \citep{Li2019}. As such, we conclude that our \nev-selected sample are not CT, in spite of their very low X-ray/\nev\ ratios.  This contrasts with findings presented in  \cite{Gilli2010} and \cite{Mignoli2013}, who argued that low X-ray/\nev\ emission in type 1 Seyferts and QSOs should be indicative of CT AGN\footnote{It may be that CT AGN have low X-ray/\nev\  ratios, as argued by \cite{Gilli2010} and \cite{Mignoli2013}.  However, the X-ray/\nev\ ratios of the \nev\ galaxies in our CLEAR sample imply that low X-ray/\nev\ ratios would then be a necessary but not sufficient condition for CT AGN}.   It therefore seems unlikely that viewing angle combined with anisotropic emission can by itself explain the enhanced \nev\ in our sample. However, we note that there is an important difference in the selection methods of our sample and the \cite{Gilli2010} and \cite{Mignoli2013} samples, where these works are at lower redshift than this work and have shallower X-ray data (100-200 ks) than the Chandra Deep Fields (2-7 Ms). These selection effects will lead to higher X-ray/\nev\ ratios in \cite{Gilli2010} and \cite{Mignoli2013}, as these samples will be insensitive to lower X-ray luminosities; e.g., the sample in \cite{Mignoli2013} reaches a flux limit of $7.3\times10^{-16}$erg s$^{-1}$cm$^{-2}$, where the CDF catalogs reach an order of magnitude fainter ($2.7\times10^{-17}$erg s$^{-1}$cm$^{-2}$ in CDF-S).
These differences in sample selection may account for some of the X-ray to \nev\ ratio discrepancies found in this work. It will be informative to study these potential selection biases in both samples with future studies of larger samples of \nev-detected objects in deep X-ray surveys.

It is still possible that the geometry of the accretion disk itself is able to produce the conditions for enhanced \nev\ emission.  This could result from an AGN with an excess of emission from the inner disk (to produce the soft X-ray photons) but less coronal emission. If this latter case applies to the galaxies in our sample of \nev\ galaxies, then it predicts we should observe SEDs with exceptionally bright UV and soft X-rays (i.e., a prominent soft X-ray excess) \citep[see, e.g.,][]{Done2012}. We may test this in future work with rest-frame far UV spectroscopy of these galaxies to more precisely constrain their hard UV/soft X-ray spectra.

The geometry may also manifest itself in the form of absorption from a warm wind.   AGN with a ``soft excess'' (of X-ray photons around $0.1$~keV) have been observed in samples of Seyferts and QSOs \citep{Walter1993}.  The origin of this emission is unclear, as the shape of the spectral energy distribution is not consistent with models of pure optically thin nor thick accretion disks \citep{Walter1993,Gierlinski2004}.   One explanation for the soft excess that is tied to the geometry is that the excess is an artifact of absorption of highly ionized atoms (e.g., \ion{O}{6}, \ion{O}{7}, and iron) in a warm, relativistic wind ejected from the accretion disk which preferentially absorbs $\sim 1$~keV photons \citep{Gierlinski2004}.     The soft excess could provide the number density of $\sim$0.1 keV photons to power the \nev\ emission in our sample.   If this is the case, we would expect to possibly see a correlation between \nev\ emission lines and broad absorption features in X-ray spectra, or with the spectral shape of the X-ray emission from $\sim$0.1 -- 50~keV data.  Currently these observations are beyond the sensitivity of X-ray telescopes. 

Regardless, our results add evidence that there is a greater diversity and variation in the intrinsic X-ray/\nev\ ratio given the complexities of the relationship between the NLR and X-ray emission. We will be able to explore these models more deeply with larger samples from \romanlong\ and greater wavelength coverage from \jwst. A full suite of UV/optical emission features in conjunction with mid-IR photometry will give a more complete picture of the physical mechanisms of these extreme high-ionization systems, and with coverage to much higher redshifts ($6<z<11$ with \jwst/NIRCam and NIRSpec). 

\section{Summary and Conclusions}\label{sec:conclusions}
In this work, we used \hst\ G102 and G141 grism observations to study a sample of 25 galaxies in the CLEAR survey displaying significant \nev\ $\lambda$3426 \AA\ emission at redshift $1.40<z<2.29$. We consider these objects of interest due to the extremely high energy (97.11 eV) required to create \nev\ compared to other strong UV/optical emission lines. Our sample selection required \snr>3 of the stronger \nev\ line (3426\AA) and \oiii\ and minimal contamination in the 1D and 2D spectra by visual inspection.

The primary findings of this work are as follows:
\begin{itemize}
    \item Galaxies with \nev\ detections are much more likely to be X-ray AGN than the general population of galaxies in the CLEAR survey. We cross-matched our sample of \nev-emitting galaxies in CLEAR with the deep (2 and 7~Ms) catalogs from the \textit{Chandra X-ray Observatory} from \cite{Luo2017} and \cite{Xue2016}.    We find that about one third ($32\%$, 8/25) of the \nev\ detected objects in our sample are X-ray-detected AGN, compared to 6.5\% of the galaxies in the mass and redshift matched CLEAR parent sample that are \textit{un}--detected in \nev.
    
    \item We use optical emission line ratios (based primarily on \oiii/\Hb) to study the ionization of the \nev-emitting galaxies.    The three spectral classifications include the mass-excitation (MEx), ``OHNO", and ``unVO87'' diagrams, which are shown in Figure \ref{fig:agn_sf}. They show that most of the \nev\ emitters are consistent with ionization with AGN, with the most reliable of these (OHNO) classifying 89\% of \nev\ galaxies as AGN. This is particularly true for X-ray-detected \nev\ sources, where all X-ray and \nev\ sources are consistent with AGN. In this work, we also include an updated redshift dependence of the MEx diagnostic, which we quantify in Equation \ref{eq:mex_z} as a shift in mass from the local \cite{Juneau2014} relation.
    
    \item There are several \nev-emitting galaxies which are not classified as AGN by X-ray or IR emission or by emission-line ratio diagnostics in Figure \ref{fig:agn_sf}. These are mostly at lower stellar masses ($\log M_\ast/M_\odot < 10$) and suggests that \nev\ selections probe AGN at intermediate mass scales or that other highly energetic photoionization mechanisms or shocks are driving the line emission.     
    \item We explore (and reject) the possibility that the \nev-emitters in our sample are produced by heavily obscured AGN by studying the X-ray/\nev\ luminosity ratio.  We find that the X-ray/\nev\ emission for our X-ray detected \nev-emitters (and  upper limits for galaxies undetected in X-rays) cannot be used to diagnose Compton thick (CT) AGN for our objects. The hydrogen absorption column densities for our objects from \cite{Li2019} support that objects in our sample are not in the CT regime $\log N_H > 24$. The use of the X-ray/\nev\ ratio to select CT AGN seems restricted to more luminous objects, such as X-ray selected QSOs and unobscured (type 1) Seyferts, which have much higher intrinsic X-ray luminosities than our \nev-selected sample.
    
    \item We argue that the \nev\ emission in our sample provides evidence for increased variation and diversity in the nature of the accretion disk and NLR of AGN at $z > 1$.  To account for the enhanced \nev\ requires an excess of ``soft X-ray'' / ``hard UV'' photons (at energies around $\sim$0.1 keV, the energy required to produce \nev). This could be a related to the ``soft excess'' seen in the spectra of other QSOs and AGN.  It could also be related to changes in the geometry, or possibly from absorption of moderately ionized gas in a relativistic wind blown off from the accretion disk.  These models can be tested by studying the spectral energy distribution of the X-ray emission, and/or by studying additional line ratios to better trace the ionizing spectrum, which should be possible with studies from, e.g., \jwst. 
\end{itemize}

Our results show that \nev\ emission probes highly energetic photoionization ($\sim$100 eV). We attribute \nev\ production predominantly to AGN activity, and we use \nev\ to probe AGN missed by other methods (X-ray and IR). Other potential creation mechanisms not explored in this work, which will be explored in future studies, include energetic shocks from supernovae and extreme ionizing stellar populations. 

Our results motivate future observations of \nev\ emission to measure the excitation of galaxies within a much larger redshift range, including the epoch of reionization ($z\gtrsim 6$). The \jwstlong\ (\jwst) will reach a flux limit that is an order of magnitude fainter than our CLEAR data for similar exposure times, enabling detection of fainter \nev-line emission. \jwst/MIRI will be particularly beneficial in the detection of 15-30 \micron\ emission from the AGN in our sample to resolve the geometry of the X-ray anisotropy. \jwst\ is outfitted with NIRSpec and NIRCam, which will give both slit and slitless spectroscopy covering strong  UV high-ionization emission lines, like the \nev\ doublet, at $0.8<z<14.4$. \jwst/NIRISS will also give slitless coverage of \nev\ at slightly lower redshift ranges ($3<z<7$).

The \romanlong\ will also be able to study the spectra of high-ionization systems in samples orders of magnitude larger than any previous work. With the Wide Field Instrument (WFI), \textit{Roman} will give low-resolution (R$\sim$600) multi-object slitless grism spectroscopy with wavelength coverage 1-1.93 \micron, similar to that of \hst/WFC3 G102+G141 but with a field of view two orders of magnitude larger in area. 

Lastly, we note that first-look \jwst/NIRSpec spectra have already shown strong detections of UV and optical spectral features in this redshift range \citep{Trump2023,Katz2023,Brinchmann2022,Cleri2023}. Early results show great promise that this new generation of spectroscopic data will give critical insight into the nature of galaxies in the early Universe, and may decisively answer questions about the key contributors to the epoch of reionization.


\software{\texttt{grizli} \citep{Brammer2008}, FAST \citep{Kriek2009}, EAZY \citep[]{Brammer2008, Wuyts2011}, Astropy \citep{Astropy2013}, NumPy \cite{Harris2020}, Matplotlib \citep{Hunter2007}}, \texttt{PyNeb} \citep{Luridiana2015}, seaborn \citep{Waskom2021}, pandas \citep{Reback2022}

\acknowledgements 

The authors wish to thank our colleagues in the CLEAR collaboration for their work on this project, and their assistance and support. NJC also thanks Maeve Curliss for significant discussion on the data visualization in this work. NJC also thanks Justin Spilker (Justin with the PhD) and Justin Cole (Justin without the PhD) for insightful discussions throughout the course of this work. NJC also acknowledges the rejected acronym for AGN exhibiting strong UV/optical features with X-ray emission weaker/comparable to local Seyferts (\textbf{S}trong UV/\textbf{O}ptical emission-line, \textbf{N}ormal \textbf{X}-ray (STONX) AGN).

This work is based on data obtained from the Hubble Space Telescope through program number GO-14227. Support for Program number GO-14227 was provided by NASA through a grant from the Space Telescope Science Institute, which is operated by the Association of Universities for Research in Astronomy, Incorporated, under NASA contract NAS5-26555. NJC, JRT, and BEB acknowledge support from NSF grant CAREER-1945546 and NASA grants JWST-ERS-01345 and 18-2ADAP18-0177. NJC and CP also acknowledge support from NASA/\textit{HST} AR 16609.  This work acknowledges support from the NASA/ESA/CSA James Webb Space Telescope through the Space Telescope Science Institute, which is operated by the Association of Universities for Research in Astronomy, Incorporated, under NASA contract NAS5-03127. Support for program No. JWST-ERS01345 was provided through a grant from the STScI under NASA contract NAS5-03127.
\clearpage
\bibliography{library}{}

\begin{thebibliography}{}
\expandafter\ifx\csname natexlab\endcsname\relax\def\natexlab#1{#1}\fi
\providecommand{\url}[1]{\href{#1}{#1}}

\bibitem[{{Astropy Collaboration} {et~al.}(2013){Astropy Collaboration},
  {Robitaille}, {Tollerud}, {Greenfield}, {Droettboom}, {Bray}, {Aldcroft},
  {Davis}, {Ginsburg}, {Price-Whelan}, {Kerzendorf}, {Conley}, {Crighton},
  {Barbary}, {Muna}, {Ferguson}, {Grollier}, {Parikh}, {Nair}, {Unther},
  {Deil}, {Woillez}, {Conseil}, {Kramer}, {Turner}, {Singer}, {Fox}, {Weaver},
  {Zabalza}, {Edwards}, {Azalee Bostroem}, {Burke}, {Casey}, {Crawford},
  {Dencheva}, {Ely}, {Jenness}, {Labrie}, {Lim}, {Pierfederici}, {Pontzen},
  {Ptak}, {Refsdal}, {Servillat}, \& {Streicher}}]{Astropy2013}
{Astropy Collaboration}, {Robitaille}, T.~P., {Tollerud}, E.~J., {et~al.} 2013,
  \aap, 558, A33

\bibitem[{{Backhaus} {et~al.}(2022{\natexlab{a}}){Backhaus}, {Trump}, {Cleri},
  {Simons}, {Momcheva}, {Papovich}, {Estrada-Carpenter}, {Finkelstein},
  {Matharu}, {Ji}, {Weiner}, {Giavalisco}, \& {Jung}}]{Backhaus2022a}
{Backhaus}, B.~E., {Trump}, J.~R., {Cleri}, N.~J., {et~al.} 2022{\natexlab{a}},
  \apj, 926, 161

\bibitem[{{Backhaus} {et~al.}(2022{\natexlab{b}}){Backhaus}, {Bridge}, {Trump},
  {Cleri}, {Papovich}, {Simons}, {Momcheva}, {Holwerda}, {Ji}, {Jung}, \&
  {Matharu}}]{Backhaus2022b}
{Backhaus}, B.~E., {Bridge}, J.~S., {Trump}, J.~R., {et~al.}
  2022{\natexlab{b}}, arXiv e-prints, arXiv:2207.11265

\bibitem[{{Barro} {et~al.}(2019){Barro}, {P{\'e}rez-Gonz{\'a}lez}, {Cava},
  {Brammer}, {Pandya}, {Eliche Moral}, {Esquej}, {Dom{\'\i}nguez-S{\'a}nchez},
  {Alcalde Pampliega}, {Guo}, {Koekemoer}, {Trump}, {Ashby}, {Cardiel},
  {Castellano}, {Conselice}, {Dickinson}, {Dolch}, {Donley}, {Espino Briones},
  {Faber}, {Fazio}, {Ferguson}, {Finkelstein}, {Fontana}, {Galametz},
  {Gardner}, {Gawiser}, {Giavalisco}, {Grazian}, {Grogin}, {Hathi}, {Hemmati},
  {Hern{\'a}n-Caballero}, {Kocevski}, {Koo}, {Kodra}, {Lee}, {Lin}, {Lucas},
  {Mobasher}, {McGrath}, {Nandra}, {Nayyeri}, {Newman}, {Pforr}, {Peth},
  {Rafelski}, {Rodr{\'\i}guez-Munoz}, {Salvato}, {Stefanon}, {van der Wel},
  {Willner}, {Wiklind}, \& {Wuyts}}]{Barro2019}
{Barro}, G., {P{\'e}rez-Gonz{\'a}lez}, P.~G., {Cava}, A., {et~al.} 2019, \apjs,
  243, 22

\bibitem[{{Berg} {et~al.}(2019){Berg}, {Chisholm}, {Erb}, {Pogge}, {Henry}, \&
  {Olivier}}]{Berg2019}
{Berg}, D.~A., {Chisholm}, J., {Erb}, D.~K., {et~al.} 2019, \apjl, 878, L3

\bibitem[{{Berg} {et~al.}(2021){Berg}, {Chisholm}, {Erb}, {Skillman}, {Pogge},
  \& {Olivier}}]{Berg2021}
---. 2021, \apj, 922, 170

\bibitem[{{Brammer} {et~al.}(2008){Brammer}, {van Dokkum}, \&
  {Coppi}}]{Brammer2008}
{Brammer}, G.~B., {van Dokkum}, P.~G., \& {Coppi}, P. 2008, \apj, 686, 1503

\bibitem[{{Brinchmann}(2022)}]{Brinchmann2022}
{Brinchmann}, J. 2022, arXiv e-prints, arXiv:2208.07467

\bibitem[{{Bruzual} \& {Charlot}(2003)}]{Bruzual2003}
{Bruzual}, G., \& {Charlot}, S. 2003, \mnras, 344, 1000

\bibitem[{{Chabrier}(2003)}]{Chabrier2003}
{Chabrier}, G. 2003, Publications of the Astronomical Society of the Pacific,
  115, 763

\bibitem[{{Cleri} {et~al.}(2022){Cleri}, {Trump}, {Backhaus}, {Momcheva},
  {Papovich}, {Simons}, {Weiner}, {Estrada-Carpenter}, {Finkelstein},
  {Giavalisco}, {Ji}, {Jung}, {Matharu}, {Martinez}, \& {Sturm}}]{Cleri2022}
{Cleri}, N.~J., {Trump}, J.~R., {Backhaus}, B.~E., {et~al.} 2022, \apj, 929, 3

\bibitem[{{Cleri} {et~al.}(2023){Cleri}, {Olivier}, {Hutchison}, {Papovich},
  {Trump}, {Amorin}, {Backhaus}, {Berg}, {Fernandez}, {Finkelstein},
  {Fujimoto}, {Hirschmann}, {Kartaltepe}, {Kocevski}, {Simons}, {Wilkins}, \&
  {Yung}}]{Cleri2023}
{Cleri}, N.~J., {Olivier}, G.~M., {Hutchison}, T.~A., {et~al.} 2023, arXiv
  e-prints, arXiv:2301.07745

\bibitem[{{Coil} {et~al.}(2015){Coil}, {Aird}, {Reddy}, {Shapley}, {Kriek},
  {Siana}, {Mobasher}, {Freeman}, {Price}, \& {Shivaei}}]{Coil2015}
{Coil}, A.~L., {Aird}, J., {Reddy}, N., {et~al.} 2015, \apj, 801, 35

\bibitem[{{Done} {et~al.}(2012){Done}, {Davis}, {Jin}, {Blaes}, \&
  {Ward}}]{Done2012}
{Done}, C., {Davis}, S.~W., {Jin}, C., {Blaes}, O., \& {Ward}, M. 2012, \mnras,
  420, 1848

\bibitem[{{Donley} {et~al.}(2012){Donley}, {Koekemoer}, {Brusa}, {Capak},
  {Cardamone}, {Civano}, {Ilbert}, {Impey}, {Kartaltepe}, {Miyaji}, {Salvato},
  {Sanders}, {Trump}, \& {Zamorani}}]{Donley2012}
{Donley}, J.~L., {Koekemoer}, A.~M., {Brusa}, M., {et~al.} 2012, \apj, 748, 142

\bibitem[{{Erb} {et~al.}(2006){Erb}, {Shapley}, {Pettini}, {Steidel}, {Reddy},
  \& {Adelberger}}]{Erb2006}
{Erb}, D.~K., {Shapley}, A.~E., {Pettini}, M., {et~al.} 2006, \apj, 644, 813

\bibitem[{{Estrada-Carpenter} {et~al.}(2019){Estrada-Carpenter}, {Papovich},
  {Momcheva}, {Brammer}, {Long}, {Quadri}, {Bridge}, {Dickinson}, {Ferguson},
  \& {Finkelstein}}]{Estrada-Carpenter2019}
{Estrada-Carpenter}, V., {Papovich}, C., {Momcheva}, I., {et~al.} 2019, \apj,
  870, 133

\bibitem[{{Estrada-Carpenter} {et~al.}(2020){Estrada-Carpenter}, {Papovich},
  {Momcheva}, {Brammer}, {Simons}, {Bridge}, {Cleri}, {Ferguson},
  {Finkelstein}, {Giavalisco}, {Jung}, {Matharu}, {Trump}, \&
  {Weiner}}]{Estrada-Carpenter2020}
---. 2020, \apj, 898, 171

\bibitem[{{Fazio} {et~al.}(2004){Fazio}, {Hora}, {Allen}, {Ashby}, {Barmby},
  {Deutsch}, {Huang}, {Kleiner}, {Marengo}, {Megeath}, {Melnick}, {Pahre},
  {Patten}, {Polizotti}, {Smith}, {Taylor}, {Wang}, {Willner}, {Hoffmann},
  {Pipher}, {Forrest}, {McMurty}, {McCreight}, {McKelvey}, {McMurray}, {Koch},
  {Moseley}, {Arendt}, {Mentzell}, {Marx}, {Losch}, {Mayman}, {Eichhorn},
  {Krebs}, {Jhabvala}, {Gezari}, {Fixsen}, {Flores}, {Shakoorzadeh}, {Jungo},
  {Hakun}, {Workman}, {Karpati}, {Kichak}, {Whitley}, {Mann}, {Tollestrup},
  {Eisenhardt}, {Stern}, {Gorjian}, {Bhattacharya}, {Carey}, {Nelson},
  {Glaccum}, {Lacy}, {Lowrance}, {Laine}, {Reach}, {Stauffer}, {Surace},
  {Wilson}, {Wright}, {Hoffman}, {Domingo}, \& {Cohen}}]{Fazio2004}
{Fazio}, G.~G., {Hora}, J.~L., {Allen}, L.~E., {et~al.} 2004, \apjs, 154, 10

\bibitem[{{Gierli{\'n}ski} \& {Done}(2004)}]{Gierlinski2004}
{Gierli{\'n}ski}, M., \& {Done}, C. 2004, \mnras, 349, L7

\bibitem[{{Gilli} {et~al.}(2010){Gilli}, {Vignali}, {Mignoli}, {Iwasawa},
  {Comastri}, \& {Zamorani}}]{Gilli2010}
{Gilli}, R., {Vignali}, C., {Mignoli}, M., {et~al.} 2010, \aap, 519, A92

\bibitem[{{Greene} {et~al.}(2020){Greene}, {Strader}, \& {Ho}}]{Greene2020}
{Greene}, J.~E., {Strader}, J., \& {Ho}, L.~C. 2020, \araa, 58, 257

\bibitem[{{Grogin} {et~al.}(2011){Grogin}, {Kocevski}, {Faber}, {Ferguson},
  {Koekemoer}, {Riess}, {Acquaviva}, {Alexander}, {Almaini}, {Ashby}, {Barden},
  {Bell}, {Bournaud}, {Brown}, {Caputi}, {Casertano}, {Cassata}, {Castellano},
  {Challis}, {Chary}, {Cheung}, {Cirasuolo}, {Conselice}, {Roshan Cooray},
  {Croton}, {Daddi}, {Dahlen}, {Dav{\'e}}, {de Mello}, {Dekel}, {Dickinson},
  {Dolch}, {Donley}, {Dunlop}, {Dutton}, {Elbaz}, {Fazio}, {Filippenko},
  {Finkelstein}, {Fontana}, {Gardner}, {Garnavich}, {Gawiser}, {Giavalisco},
  {Grazian}, {Guo}, {Hathi}, {H{\"a}ussler}, {Hopkins}, {Huang}, {Huang},
  {Jha}, {Kartaltepe}, {Kirshner}, {Koo}, {Lai}, {Lee}, {Li}, {Lotz}, {Lucas},
  {Madau}, {McCarthy}, {McGrath}, {McIntosh}, {McLure}, {Mobasher},
  {Moustakas}, {Mozena}, {Nandra}, {Newman}, {Niemi}, {Noeske}, {Papovich},
  {Pentericci}, {Pope}, {Primack}, {Rajan}, {Ravindranath}, {Reddy}, {Renzini},
  {Rix}, {Robaina}, {Rodney}, {Rosario}, {Rosati}, {Salimbeni}, {Scarlata},
  {Siana}, {Simard}, {Smidt}, {Somerville}, {Spinrad}, {Straughn}, {Strolger},
  {Telford}, {Teplitz}, {Trump}, {van der Wel}, {Villforth}, {Wechsler},
  {Weiner}, {Wiklind}, {Wild}, {Wilson}, {Wuyts}, {Yan}, \& {Yun}}]{Grogin2011}
{Grogin}, N.~A., {Kocevski}, D.~D., {Faber}, S.~M., {et~al.} 2011, \apjs, 197,
  35

\bibitem[{{Guo} {et~al.}(2013){Guo}, {Ferguson}, {Giavalisco}, {Barro},
  {Willner}, {Ashby}, {Dahlen}, {Donley}, {Faber}, {Fontana}, {Galametz},
  {Grazian}, {Huang}, {Kocevski}, {Koekemoer}, {Koo}, {McGrath}, {Peth},
  {Salvato}, {Wuyts}, {Castellano}, {Cooray}, {Dickinson}, {Dunlop}, {Fazio},
  {Gardner}, {Gawiser}, {Grogin}, {Hathi}, {Hsu}, {Lee}, {Lucas}, {Mobasher},
  {Nandra}, {Newman}, \& {van der Wel}}]{Guo2013}
{Guo}, Y., {Ferguson}, H.~C., {Giavalisco}, M., {et~al.} 2013, \apjs, 207, 24

\bibitem[{Harris {et~al.}(2020)Harris, Millman, van~der Walt, Gommers,
  Virtanen, Cournapeau, Wieser, Taylor, Berg, Smith, Kern, Picus, Hoyer, van
  Kerkwijk, Brett, Haldane, del R{\'{i}}o, Wiebe, Peterson,
  G{\'{e}}rard-Marchant, Sheppard, Reddy, Weckesser, Abbasi, Gohlke, \&
  Oliphant}]{Harris2020}
Harris, C.~R., Millman, K.~J., van~der Walt, S.~J., {et~al.} 2020, Nature, 585,
  357.
\newblock \url{https://doi.org/10.1038/s41586-020-2649-2}

\bibitem[{{Heckman} {et~al.}(2005){Heckman}, {Ptak}, {Hornschemeier}, \&
  {Kauffmann}}]{Heckman2005}
{Heckman}, T.~M., {Ptak}, A., {Hornschemeier}, A., \& {Kauffmann}, G. 2005,
  \apj, 634, 161

\bibitem[{{Hickox} \& {Alexander}(2018)}]{Hickox2018}
{Hickox}, R.~C., \& {Alexander}, D.~M. 2018, \araa, 56, 625

\bibitem[{{Hunter}(2007)}]{Hunter2007}
{Hunter}, J.~D. 2007, Computing in Science and Engineering, 9, 90

\bibitem[{{Izotov} {et~al.}(2021){Izotov}, {Thuan}, \& {Guseva}}]{Izotov2021}
{Izotov}, Y.~I., {Thuan}, T.~X., \& {Guseva}, N.~G. 2021, \mnras, 508, 2556

\bibitem[{{Izotov} {et~al.}(2012){Izotov}, {Thuan}, \& {Privon}}]{Izotov2012}
{Izotov}, Y.~I., {Thuan}, T.~X., \& {Privon}, G. 2012, \mnras, 427, 1229

\bibitem[{{Juneau} {et~al.}(2011){Juneau}, {Dickinson}, {Alexander}, \&
  {Salim}}]{Juneau2011}
{Juneau}, S., {Dickinson}, M., {Alexander}, D.~M., \& {Salim}, S. 2011, \apj,
  736, 104

\bibitem[{{Juneau} {et~al.}(2014){Juneau}, {Bournaud}, {Charlot}, {Daddi},
  {Elbaz}, {Trump}, {Brinchmann}, {Dickinson}, {Duc}, {Gobat}, {Jean-Baptiste},
  {Le Floc'h}, {Lehnert}, {Pacifici}, {Pannella}, \& {Schreiber}}]{Juneau2014}
{Juneau}, S., {Bournaud}, F., {Charlot}, S., {et~al.} 2014, \apj, 788, 88

\bibitem[{{Jung} {et~al.}(2022){Jung}, {Papovich}, {Finkelstein}, {Simons},
  {Estrada-Carpenter}, {Backhaus}, {Cleri}, {Finlator}, {Giavalisco}, {Ji},
  {Matharu}, {Momcheva}, {Straughn}, \& {Trump}}]{Jung2022}
{Jung}, I., {Papovich}, C., {Finkelstein}, S.~L., {et~al.} 2022, \apj, 933, 87

\bibitem[{{Katz} {et~al.}(2023){Katz}, {Saxena}, {Cameron}, {Carniani},
  {Bunker}, {Arribas}, {Bhatawdekar}, {Bowler}, {Boyett}, {Cresci},
  {Curtis-Lake}, {D'Eugenio}, {Kumari}, {Looser}, {Maiolino}, {{\"U}bler},
  {Willott}, \& {Witstok}}]{Katz2023}
{Katz}, H., {Saxena}, A., {Cameron}, A.~J., {et~al.} 2023, \mnras, 518, 592

\bibitem[{{Kauffmann} {et~al.}(2003){Kauffmann}, {Heckman}, {Tremonti},
  {Brinchmann}, {Charlot}, {White}, {Ridgway}, {Brinkmann}, {Fukugita}, {Hall},
  {Ivezi{\'c}}, {Richards}, \& {Schneider}}]{Kauffmann2003}
{Kauffmann}, G., {Heckman}, T.~M., {Tremonti}, C., {et~al.} 2003, \mnras, 346,
  1055

\bibitem[{{Kennicutt}(1998)}]{Kennicutt1998}
{Kennicutt}, Robert~C., J. 1998, \apj, 498, 541

\bibitem[{{Kewley} {et~al.}(2001){Kewley}, {Heisler}, {Dopita}, \&
  {Lumsden}}]{Kewley2001}
{Kewley}, L.~J., {Heisler}, C.~A., {Dopita}, M.~A., \& {Lumsden}, S. 2001,
  \apjs, 132, 37

\bibitem[{{Kewley} {et~al.}(2019{\natexlab{a}}){Kewley}, {Nicholls},
  {Sutherland}, {Rigby}, {Acharya}, {Dopita}, \& {Bayliss}}]{Kewley2019a}
{Kewley}, L.~J., {Nicholls}, D.~C., {Sutherland}, R., {et~al.}
  2019{\natexlab{a}}, \apj, 880, 16

\bibitem[{{Kewley} {et~al.}(2019{\natexlab{b}}){Kewley}, {Nicholls}, \&
  {Sutherland}}]{Kewley2019b}
{Kewley}, L.~J., {Nicholls}, D.~C., \& {Sutherland}, R.~S. 2019{\natexlab{b}},
  \araa, 57, 511

\bibitem[{{Koekemoer} {et~al.}(2011){Koekemoer}, {Faber}, {Ferguson}, {Grogin},
  {Kocevski}, {Koo}, {Lai}, {Lotz}, {Lucas}, {McGrath}, {Ogaz}, {Rajan},
  {Riess}, {Rodney}, {Strolger}, {Casertano}, {Castellano}, {Dahlen},
  {Dickinson}, {Dolch}, {Fontana}, {Giavalisco}, {Grazian}, {Guo}, {Hathi},
  {Huang}, {van der Wel}, {Yan}, {Acquaviva}, {Alexander}, {Almaini}, {Ashby},
  {Barden}, {Bell}, {Bournaud}, {Brown}, {Caputi}, {Cassata}, {Challis},
  {Chary}, {Cheung}, {Cirasuolo}, {Conselice}, {Roshan Cooray}, {Croton},
  {Daddi}, {Dav{\'e}}, {de Mello}, {de Ravel}, {Dekel}, {Donley}, {Dunlop},
  {Dutton}, {Elbaz}, {Fazio}, {Filippenko}, {Finkelstein}, {Frazer}, {Gardner},
  {Garnavich}, {Gawiser}, {Gruetzbauch}, {Hartley}, {H{\"a}ussler},
  {Herrington}, {Hopkins}, {Huang}, {Jha}, {Johnson}, {Kartaltepe},
  {Khostovan}, {Kirshner}, {Lani}, {Lee}, {Li}, {Madau}, {McCarthy},
  {McIntosh}, {McLure}, {McPartland}, {Mobasher}, {Moreira}, {Mortlock},
  {Moustakas}, {Mozena}, {Nandra}, {Newman}, {Nielsen}, {Niemi}, {Noeske},
  {Papovich}, {Pentericci}, {Pope}, {Primack}, {Ravindranath}, {Reddy},
  {Renzini}, {Rix}, {Robaina}, {Rosario}, {Rosati}, {Salimbeni}, {Scarlata},
  {Siana}, {Simard}, {Smidt}, {Snyder}, {Somerville}, {Spinrad}, {Straughn},
  {Telford}, {Teplitz}, {Trump}, {Vargas}, {Villforth}, {Wagner}, {Wandro},
  {Wechsler}, {Weiner}, {Wiklind}, {Wild}, {Wilson}, {Wuyts}, \&
  {Yun}}]{Koekemoer2011}
{Koekemoer}, A.~M., {Faber}, S.~M., {Ferguson}, H.~C., {et~al.} 2011, \apjs,
  197, 36

\bibitem[{{Kriek} {et~al.}(2009){Kriek}, {van Dokkum}, {Labb{\'e}}, {Franx},
  {Illingworth}, {Marchesini}, \& {Quadri}}]{Kriek2009}
{Kriek}, M., {van Dokkum}, P.~G., {Labb{\'e}}, I., {et~al.} 2009, \apj, 700,
  221

\bibitem[{{Lacy} {et~al.}(2004){Lacy}, {Storrie-Lombardi}, {Sajina},
  {Appleton}, {Armus}, {Chapman}, {Choi}, {Fadda}, {Fang}, {Frayer},
  {Heinrichsen}, {Helou}, {Im}, {Marleau}, {Masci}, {Shupe}, {Soifer},
  {Surace}, {Teplitz}, {Wilson}, \& {Yan}}]{Lacy2004}
{Lacy}, M., {Storrie-Lombardi}, L.~J., {Sajina}, A., {et~al.} 2004, \apjs, 154,
  166

\bibitem[{{Lambrides} {et~al.}(2020){Lambrides}, {Chiaberge}, {Heckman},
  {Gilli}, {Vito}, \& {Norman}}]{Lambrides2020}
{Lambrides}, E.~L., {Chiaberge}, M., {Heckman}, T., {et~al.} 2020, \apj, 897,
  160

\bibitem[{{Lee} {et~al.}(2014){Lee}, {Pirzkal}, \& {Hilbert}}]{Lee2014}
{Lee}, J.~C., {Pirzkal}, N., \& {Hilbert}, B. 2014, {Flux Calibration
  Monitoring: WFC3/IR G102 and G141 Grisms}, Tech. rep.

\bibitem[{{Lehmer} {et~al.}(2010){Lehmer}, {Alexander}, {Bauer}, {Brandt},
  {Goulding}, {Jenkins}, {Ptak}, \& {Roberts}}]{Lehmer2010}
{Lehmer}, B.~D., {Alexander}, D.~M., {Bauer}, F.~E., {et~al.} 2010, \apj, 724,
  559

\bibitem[{{Leung} {et~al.}(2021){Leung}, {Coil}, {Rupke}, \&
  {Perrotta}}]{Leung2021}
{Leung}, G. C.~K., {Coil}, A.~L., {Rupke}, D. S.~N., \& {Perrotta}, S. 2021,
  \apj, 914, 17

\bibitem[{{Levesque} \& {Richardson}(2014)}]{Levesque2014}
{Levesque}, E.~M., \& {Richardson}, M. L.~A. 2014, \apj, 780, 100

\bibitem[{{Li} {et~al.}(2019){Li}, {Xue}, {Sun}, {Liu}, {Vito}, {Brandt},
  {Hughes}, {Yang}, {Tozzi}, {Zhu}, {Zheng}, {Luo}, {Chen}, {Vignali}, {Gilli},
  \& {Shu}}]{Li2019}
{Li}, J., {Xue}, Y., {Sun}, M., {et~al.} 2019, \apj, 877, 5

\bibitem[{{Liu} {et~al.}(2017){Liu}, {Tozzi}, {Wang}, {Brandt}, {Vignali},
  {Xue}, {Schneider}, {Comastri}, {Yang}, {Bauer}, {Paolillo}, {Luo}, {Gilli},
  {Wang}, {Giavalisco}, {Ji}, {Alexander}, {Mainieri}, {Shemmer}, {Koekemoer},
  \& {Risaliti}}]{Liu2017}
{Liu}, T., {Tozzi}, P., {Wang}, J.-X., {et~al.} 2017, \apjs, 232, 8

\bibitem[{{Luo} {et~al.}(2017){Luo}, {Brandt}, {Xue}, {Lehmer}, {Alexander},
  {Bauer}, {Vito}, {Yang}, {Basu-Zych}, {Comastri}, {Gilli}, {Gu},
  {Hornschemeier}, {Koekemoer}, {Liu}, {Mainieri}, {Paolillo}, {Ranalli},
  {Rosati}, {Schneider}, {Shemmer}, {Smail}, {Sun}, {Tozzi}, {Vignali}, \&
  {Wang}}]{Luo2017}
{Luo}, B., {Brandt}, W.~N., {Xue}, Y.~Q., {et~al.} 2017, The Astrophysical
  Journal Supplement Series, 228, 2

\bibitem[{{Luridiana} {et~al.}(2015){Luridiana}, {Morisset}, \&
  {Shaw}}]{Luridiana2015}
{Luridiana}, V., {Morisset}, C., \& {Shaw}, R.~A. 2015, \aap, 573, A42

\bibitem[{{Madau} \& {Dickinson}(2014)}]{Madau2014}
{Madau}, P., \& {Dickinson}, M. 2014, \araa, 52, 415

\bibitem[{{Maiolino} {et~al.}(1998){Maiolino}, {Salvati}, {Bassani}, {Dadina},
  {della Ceca}, {Matt}, {Risaliti}, \& {Zamorani}}]{Maiolino1998}
{Maiolino}, R., {Salvati}, M., {Bassani}, L., {et~al.} 1998, \aap, 338, 781

\bibitem[{{Masters} {et~al.}(2014){Masters}, {McCarthy}, {Siana}, {Malkan},
  {Mobasher}, {Atek}, {Henry}, {Martin}, {Rafelski}, {Hathi}, {Scarlata},
  {Ross}, {Bunker}, {Blanc}, {Bedregal}, {Dom{\'\i}nguez}, {Colbert},
  {Teplitz}, \& {Dressler}}]{Masters2014}
{Masters}, D., {McCarthy}, P., {Siana}, B., {et~al.} 2014, \apj, 785, 153

\bibitem[{{Matharu} {et~al.}(2022){Matharu}, {Papovich}, {Simons}, {Momcheva},
  {Brammer}, {Ji}, {Backhaus}, {Cleri}, {Estrada-Carpenter}, {Finkelstein},
  {Finlator}, {Giavalisco}, {Jung}, {Muzzin}, {Pillepich}, {Trump}, \&
  {Weiner}}]{Matharu2022}
{Matharu}, J., {Papovich}, C., {Simons}, R.~C., {et~al.} 2022, arXiv e-prints,
  arXiv:2205.08543

\bibitem[{{Mendez} {et~al.}(2013){Mendez}, {Coil}, {Aird}, {Diamond-Stanic},
  {Moustakas}, {Blanton}, {Cool}, {Eisenstein}, {Wong}, \& {Zhu}}]{Mendez2013}
{Mendez}, A.~J., {Coil}, A.~L., {Aird}, J., {et~al.} 2013, \apj, 770, 40

\bibitem[{{Mignoli} {et~al.}(2013){Mignoli}, {Vignali}, {Gilli}, {Comastri},
  {Zamorani}, {Bolzonella}, {Bongiorno}, {Lamareille}, {Nair}, {Pozzetti},
  {Lilly}, {Carollo}, {Contini}, {Kneib}, {Le F{\`e}vre}, {Mainieri},
  {Renzini}, {Scodeggio}, {Bardelli}, {Caputi}, {Cucciati}, {de la Torre}, {de
  Ravel}, {Franzetti}, {Garilli}, {Iovino}, {Kampczyk}, {Knobel},
  {Kova{\v{c}}}, {Le Borgne}, {Le Brun}, {Maier}, {Pell{\`o}}, {Peng}, {Perez
  Montero}, {Presotto}, {Silverman}, {Tanaka}, {Tasca}, {Tresse}, {Vergani},
  {Zucca}, {Bordoloi}, {Cappi}, {Cimatti}, {Koekemoer}, {McCracken}, {Moresco},
  \& {Welikala}}]{Mignoli2013}
{Mignoli}, M., {Vignali}, C., {Gilli}, R., {et~al.} 2013, \aap, 556, A29

\bibitem[{{Momcheva} {et~al.}(2016){Momcheva}, {Brammer}, {van Dokkum},
  {Skelton}, {Whitaker}, {Nelson}, {Fumagalli}, {Maseda}, {Leja}, {Franx},
  {Rix}, {Bezanson}, {Da Cunha}, {Dickey}, {F{\"o}rster Schreiber},
  {Illingworth}, {Kriek}, {Labb{\'e}}, {Ulf Lange}, {Lundgren}, {Magee},
  {Marchesini}, {Oesch}, {Pacifici}, {Patel}, {Price}, {Tal}, {Wake}, {van der
  Wel}, \& {Wuyts}}]{Momcheva2016}
{Momcheva}, I.~G., {Brammer}, G.~B., {van Dokkum}, P.~G., {et~al.} 2016, \apjs,
  225, 27

\bibitem[{{Nakajima} \& {Ouchi}(2014)}]{Nakajima2014}
{Nakajima}, K., \& {Ouchi}, M. 2014, \mnras, 442, 900

\bibitem[{{Olivier} {et~al.}(2022){Olivier}, {Berg}, {Chisholm}, {Erb},
  {Pogge}, \& {Skillman}}]{Olivier2022}
{Olivier}, G.~M., {Berg}, D.~A., {Chisholm}, J., {et~al.} 2022, \apj, 938, 16

\bibitem[{{Papovich} {et~al.}(2022){Papovich}, {Simons}, {Estrada-Carpenter},
  {Matharu}, {Momcheva}, {Trump}, {Backhaus}, {Brammer}, {Cleri},
  {Finkelstein}, {Giavalisco}, {Ji}, {Jung}, {Kewley}, {Nicholls}, {Pirzkal},
  {Rafelski}, \& {Weiner}}]{Papovich2022}
{Papovich}, C., {Simons}, R.~C., {Estrada-Carpenter}, V., {et~al.} 2022, arXiv
  e-prints, arXiv:2205.05090

\bibitem[{{P{\'e}rez-Gonz{\'a}lez} {et~al.}(2013){P{\'e}rez-Gonz{\'a}lez},
  {Cava}, {Barro}, {Villar}, {Cardiel}, {Ferreras}, {Rodr{\'\i}guez-Espinosa},
  {Alonso-Herrero}, {Balcells}, {Cenarro}, {Cepa}, {Charlot}, {Cimatti},
  {Conselice}, {Daddi}, {Donley}, {Elbaz}, {Espino}, {Gallego}, {Gobat},
  {Gonz{\'a}lez-Mart{\'\i}n}, {Guzm{\'a}n}, {Hern{\'a}n-Caballero},
  {Mu{\~n}oz-Tu{\~n}{\'o}n}, {Renzini}, {Rodr{\'\i}guez-Zaur{\'\i}n}, {Tresse},
  {Trujillo}, \& {Zamorano}}]{Perez-Gonzalez2013}
{P{\'e}rez-Gonz{\'a}lez}, P.~G., {Cava}, A., {Barro}, G., {et~al.} 2013, \apj,
  762, 46

\bibitem[{{Pirzkal} {et~al.}(2016){Pirzkal}, {Ryan}, \&
  {Brammer}}]{Pirzkal2016}
{Pirzkal}, N., {Ryan}, R., \& {Brammer}, G. 2016, {Trace and Wavelength
  Calibrations of the WFC3 G102 and G141 IR Grisms}, Instrument Science Report
  WFC3 2016-15, 25 pages, ,

\bibitem[{{Pirzkal} {et~al.}(2017){Pirzkal}, {Malhotra}, {Ryan}, {Rothberg},
  {Grogin}, {Finkelstein}, {Koekemoer}, {Rhoads}, {Larson}, {Christensen},
  {Cimatti}, {Ferreras}, {Gardner}, {Gronwall}, {Hathi}, {Hibon}, {Joshi},
  {Kuntschner}, {Meurer}, {O'Connell}, {Oestlin}, {Pasquali}, {Pharo},
  {Straughn}, {Walsh}, {Watson}, {Windhorst}, {Zakamska}, \&
  {Zirm}}]{Pirzkal2017}
{Pirzkal}, N., {Malhotra}, S., {Ryan}, R.~E., {et~al.} 2017, \apj, 846, 84

\bibitem[{{Planck Collaboration} {et~al.}(2020){Planck Collaboration},
  {Aghanim}, {Akrami}, {Ashdown}, {Aumont}, {Baccigalupi}, {Ballardini},
  {Banday}, {Barreiro}, {Bartolo}, {Basak}, {Battye}, {Benabed}, {Bernard},
  {Bersanelli}, {Bielewicz}, {Bock}, {Bond}, {Borrill}, {Bouchet}, {Boulanger},
  {Bucher}, {Burigana}, {Butler}, {Calabrese}, {Cardoso}, {Carron},
  {Challinor}, {Chiang}, {Chluba}, {Colombo}, {Combet}, {Contreras}, {Crill},
  {Cuttaia}, {de Bernardis}, {de Zotti}, {Delabrouille}, {Delouis}, {Di
  Valentino}, {Diego}, {Dor{\'e}}, {Douspis}, {Ducout}, {Dupac}, {Dusini},
  {Efstathiou}, {Elsner}, {En{\ss}lin}, {Eriksen}, {Fantaye}, {Farhang},
  {Fergusson}, {Fernandez-Cobos}, {Finelli}, {Forastieri}, {Frailis},
  {Fraisse}, {Franceschi}, {Frolov}, {Galeotta}, {Galli}, {Ganga},
  {G{\'e}nova-Santos}, {Gerbino}, {Ghosh}, {Gonz{\'a}lez-Nuevo}, {G{\'o}rski},
  {Gratton}, {Gruppuso}, {Gudmundsson}, {Hamann}, {Handley}, {Hansen},
  {Herranz}, {Hildebrandt}, {Hivon}, {Huang}, {Jaffe}, {Jones}, {Karakci},
  {Keih{\"a}nen}, {Keskitalo}, {Kiiveri}, {Kim}, {Kisner}, {Knox},
  {Krachmalnicoff}, {Kunz}, {Kurki-Suonio}, {Lagache}, {Lamarre}, {Lasenby},
  {Lattanzi}, {Lawrence}, {Le Jeune}, {Lemos}, {Lesgourgues}, {Levrier},
  {Lewis}, {Liguori}, {Lilje}, {Lilley}, {Lindholm}, {L{\'o}pez-Caniego},
  {Lubin}, {Ma}, {Mac{\'\i}as-P{\'e}rez}, {Maggio}, {Maino}, {Mandolesi},
  {Mangilli}, {Marcos-Caballero}, {Maris}, {Martin}, {Martinelli},
  {Mart{\'\i}nez-Gonz{\'a}lez}, {Matarrese}, {Mauri}, {McEwen}, {Meinhold},
  {Melchiorri}, {Mennella}, {Migliaccio}, {Millea}, {Mitra},
  {Miville-Desch{\^e}nes}, {Molinari}, {Montier}, {Morgante}, {Moss}, {Natoli},
  {N{\o}rgaard-Nielsen}, {Pagano}, {Paoletti}, {Partridge}, {Patanchon},
  {Peiris}, {Perrotta}, {Pettorino}, {Piacentini}, {Polastri}, {Polenta},
  {Puget}, {Rachen}, {Reinecke}, {Remazeilles}, {Renzi}, {Rocha}, {Rosset},
  {Roudier}, {Rubi{\~n}o-Mart{\'\i}n}, {Ruiz-Granados}, {Salvati}, {Sandri},
  {Savelainen}, {Scott}, {Shellard}, {Sirignano}, {Sirri}, {Spencer},
  {Sunyaev}, {Suur-Uski}, {Tauber}, {Tavagnacco}, {Tenti}, {Toffolatti},
  {Tomasi}, {Trombetti}, {Valenziano}, {Valiviita}, {Van Tent}, {Vibert},
  {Vielva}, {Villa}, {Vittorio}, {Wandelt}, {Wehus}, {White}, {White},
  {Zacchei}, \& {Zonca}}]{Planck2020}
{Planck Collaboration}, {Aghanim}, N., {Akrami}, Y., {et~al.} 2020, \aap, 641,
  A6

\bibitem[{{Reback} {et~al.}(2022){Reback}, {jbrockmendel}, {McKinney}, {Van den
  Bossche}, {Augspurger}, {Roeschke}, {Hawkins}, {Cloud}, {gfyoung}, {Sinhrks},
  {Hoefler}, {Klein}, {Petersen}, {Tratner}, {She}, {Ayd}, {Naveh},
  {Darbyshire}, {Garcia}, {Shadrach}, {Schendel}, {Hayden}, {Saxton},
  {Gorelli}, {Li}, {Zeitlin}, {Jancauskas}, {McMaster}, {W{\"o}rtwein}, \&
  {Battiston}}]{Reback2022}
{Reback}, J., {jbrockmendel}, {McKinney}, W., {et~al.} 2022,
  {pandas-dev/pandas: Pandas 1.4.2}, Zenodo, vv1.4.2,  Zenodo,
  doi:10.5281/zenodo.3509134

\bibitem[{{Rhoads} {et~al.}(2022){Rhoads}, {Wold}, {Harish}, {Kim}, {Pharo},
  {Malhotra}, {Gabrielpillai}, {Jiang}, \& {Yang}}]{Rhoads2022}
{Rhoads}, J.~E., {Wold}, I. G.~B., {Harish}, S., {et~al.} 2022, arXiv e-prints,
  arXiv:2207.13020

\bibitem[{{Simons} {et~al.}(2021){Simons}, {Papovich}, {Momcheva}, {Trump},
  {Brammer}, {Estrada-Carpenter}, {Backhaus}, {Cleri}, {Finkelstein},
  {Giavalisco}, {Ji}, {Jung}, {Matharu}, \& {Weiner}}]{Simons2021}
{Simons}, R.~C., {Papovich}, C., {Momcheva}, I., {et~al.} 2021, \apj, 923, 203

\bibitem[{{Simons} {et~al.}(2023){Simons}, {Papovich}, {Momcheva}, {Brammer},
  {Estrada-Carpenter}, {Finkelstein}, {Gosmeyer}, {Matharu}, {Trump},
  {Backhaus}, {Cheng}, {Cleri}, {Ferguson}, {Finlator}, {Giavalisco}, {Ji},
  {Jung}, {Lotz}, {O'Brien}, {Skelton}, {Tilvi}, \& {Weiner}}]{Simons2023}
{Simons}, R.~C., {Papovich}, C., {Momcheva}, I.~G., {et~al.} 2023, arXiv
  e-prints, arXiv:2303.09570

\bibitem[{{Skelton} {et~al.}(2014){Skelton}, {Whitaker}, {Momcheva}, {Brammer},
  {van Dokkum}, {Labb{\'e}}, {Franx}, {van der Wel}, {Bezanson}, {Da Cunha},
  {Fumagalli}, {F{\"o}rster Schreiber}, {Kriek}, {Leja}, {Lundgren}, {Magee},
  {Marchesini}, {Maseda}, {Nelson}, {Oesch}, {Pacifici}, {Patel}, {Price},
  {Rix}, {Tal}, {Wake}, \& {Wuyts}}]{Skelton2014}
{Skelton}, R.~E., {Whitaker}, K.~E., {Momcheva}, I.~G., {et~al.} 2014, \apjs,
  214, 24

\bibitem[{{Steidel} {et~al.}(2014){Steidel}, {Rudie}, {Strom}, {Pettini},
  {Reddy}, {Shapley}, {Trainor}, {Erb}, {Turner}, {Konidaris}, {Kulas}, {Mace},
  {Matthews}, \& {McLean}}]{Steidel2014}
{Steidel}, C.~C., {Rudie}, G.~C., {Strom}, A.~L., {et~al.} 2014, \apj, 795, 165

\bibitem[{{Stern} {et~al.}(2005){Stern}, {Eisenhardt}, {Gorjian}, {Kochanek},
  {Caldwell}, {Eisenstein}, {Brodwin}, {Brown}, {Cool}, {Dey}, {Green},
  {Jannuzi}, {Murray}, {Pahre}, \& {Willner}}]{Stern2005}
{Stern}, D., {Eisenhardt}, P., {Gorjian}, V., {et~al.} 2005, \apj, 631, 163

\bibitem[{{Tang} {et~al.}(2021){Tang}, {Stark}, {Chevallard}, {Charlot},
  {Endsley}, \& {Congiu}}]{Tang2021}
{Tang}, M., {Stark}, D.~P., {Chevallard}, J., {et~al.} 2021, \mnras, 501, 3238

\bibitem[{{Thuan} \& {Izotov}(2005)}]{Thuan2005}
{Thuan}, T.~X., \& {Izotov}, Y.~I. 2005, \apjs, 161, 240

\bibitem[{{Trump} {et~al.}(2011){Trump}, {Impey}, {Kelly}, {Civano}, {Gabor},
  {Diamond-Stanic}, {Merloni}, {Urry}, {Hao}, {Jahnke}, {Nagao}, {Taniguchi},
  {Koekemoer}, {Lanzuisi}, {Liu}, {Mainieri}, {Salvato}, \&
  {Scoville}}]{Trump2011}
{Trump}, J.~R., {Impey}, C.~D., {Kelly}, B.~C., {et~al.} 2011, \apj, 733, 60

\bibitem[{{Trump} {et~al.}(2015){Trump}, {Sun}, {Zeimann}, {Luck}, {Bridge},
  {Grier}, {Hagen}, {Juneau}, {Montero-Dorta}, {Rosario}, {Brandt},
  {Ciardullo}, \& {Schneider}}]{Trump2015}
{Trump}, J.~R., {Sun}, M., {Zeimann}, G.~R., {et~al.} 2015, \apj, 811, 26

\bibitem[{{Trump} {et~al.}(2023){Trump}, {Arrabal Haro}, {Simons}, {Backhaus},
  {Amor{\'\i}n}, {Dickinson}, {Fern{\'a}ndez}, {Papovich}, {Nicholls},
  {Kewley}, {Brunker}, {Salzer}, {Wilkins}, {Almaini}, {Bagley}, {Berg},
  {Bhatawdekar}, {Bisigello}, {Buat}, {Burgarella}, {Calabr{\`o}}, {Casey},
  {Ciesla}, {Cleri}, {Cole}, {Cooper}, {Cooray}, {Costantin}, {Croton},
  {Ferguson}, {Finkelstein}, {Fujimoto}, {Gardner}, {Gawiser}, {Giavalisco},
  {Grazian}, {Grogin}, {Hathi}, {Hirschmann}, {Holwerda}, {Huertas-Company},
  {Hutchison}, {Jogee}, {Juneau}, {Jung}, {Kartaltepe}, {Kirkpatrick},
  {Kocevski}, {Koekemoer}, {Lotz}, {Lucas}, {Magnelli}, {Matharu},
  {P{\'e}rez-Gonz{\'a}lez}, {Pirzkal}, {Rafelski}, {Rose}, {Seill{\'e}},
  {Somerville}, {Straughn}, {Tacchella}, {Vanderhoof}, {Weiner}, {Wuyts},
  {Yung}, \& {Zavala}}]{Trump2023}
{Trump}, J.~R., {Arrabal Haro}, P., {Simons}, R.~C., {et~al.} 2023, \apj, 945,
  35

\bibitem[{{Veilleux} \& {Osterbrock}(1987)}]{Veilleux1987}
{Veilleux}, S., \& {Osterbrock}, D.~E. 1987, \apjs, 63, 295

\bibitem[{{Walter} \& {Fink}(1993)}]{Walter1993}
{Walter}, R., \& {Fink}, H.~H. 1993, \aap, 274, 105

\bibitem[{{Waskom}(2021)}]{Waskom2021}
{Waskom}, M. 2021, The Journal of Open Source Software, 6, 3021

\bibitem[{{Wuyts} {et~al.}(2011){Wuyts}, {F{\"o}rster Schreiber}, {Lutz},
  {Nordon}, {Berta}, {Altieri}, {Andreani}, {Aussel}, {Bongiovanni}, {Cepa},
  {Cimatti}, {Daddi}, {Elbaz}, {Genzel}, {Koekemoer}, {Magnelli}, {Maiolino},
  {McGrath}, {P{\'e}rez Garc{\'\i}a}, {Poglitsch}, {Popesso}, {Pozzi},
  {Sanchez-Portal}, {Sturm}, {Tacconi}, \& {Valtchanov}}]{Wuyts2011}
{Wuyts}, S., {F{\"o}rster Schreiber}, N.~M., {Lutz}, D., {et~al.} 2011, \apj,
  738, 106

\bibitem[{{Xue} {et~al.}(2016){Xue}, {Luo}, {Brandt}, {Alexander}, {Bauer},
  {Lehmer}, \& {Yang}}]{Xue2016}
{Xue}, Y.~Q., {Luo}, B., {Brandt}, W.~N., {et~al.} 2016, \apjs, 224, 15

\bibitem[{{Xue} {et~al.}(2011){Xue}, {Luo}, {Brandt}, {Bauer}, {Lehmer},
  {Broos}, {Schneider}, {Alexander}, {Brusa}, {Comastri}, {Fabian}, {Gilli},
  {Hasinger}, {Hornschemeier}, {Koekemoer}, {Liu}, {Mainieri}, {Paolillo},
  {Rafferty}, {Rosati}, {Shemmer}, {Silverman}, {Smail}, {Tozzi}, \&
  {Vignali}}]{Xue2011}
---. 2011, \apjs, 195, 10

\bibitem[{{Yan} {et~al.}(2011){Yan}, {Ho}, {Newman}, {Coil}, {Willmer},
  {Laird}, {Georgakakis}, {Aird}, {Barmby}, {Bundy}, {Cooper}, {Davis},
  {Faber}, {Fang}, {Griffith}, {Koekemoer}, {Koo}, {Nandra}, {Park},
  {Sarajedini}, {Weiner}, \& {Willner}}]{Yan2011}
{Yan}, R., {Ho}, L.~C., {Newman}, J.~A., {et~al.} 2011, \apj, 728, 38

\bibitem[{{Yang} {et~al.}(2016){Yang}, {Brandt}, {Luo}, {Xue}, {Bauer}, {Sun},
  {Kim}, {Schulze}, {Zheng}, {Paolillo}, {Shemmer}, {Liu}, {Schneider},
  {Vignali}, {Vito}, \& {Wang}}]{Yang2016}
{Yang}, G., {Brandt}, W.~N., {Luo}, B., {et~al.} 2016, \apj, 831, 145

\bibitem[{{Yang} {et~al.}(2020){Yang}, {Boquien}, {Buat}, {Burgarella},
  {Ciesla}, {Duras}, {Stalevski}, {Brandt}, \& {Papovich}}]{Yang2020}
{Yang}, G., {Boquien}, M., {Buat}, V., {et~al.} 2020, \mnras, 491, 740

\bibitem[{{Zeimann} {et~al.}(2014){Zeimann}, {Ciardullo}, {Gebhardt},
  {Gronwall}, {Schneider}, {Hagen}, {Bridge}, {Feldmeier}, \&
  {Trump}}]{Zeimann2014}
{Zeimann}, G.~R., {Ciardullo}, R., {Gebhardt}, H., {et~al.} 2014, \apj, 790,
  113

\bibitem[{{Zeimann} {et~al.}(2015){Zeimann}, {Ciardullo}, {Gebhardt},
  {Gronwall}, {Hagen}, {Trump}, {Bridge}, {Luo}, \& {Schneider}}]{Zeimann2015}
---. 2015, \apj, 798, 29

\end{thebibliography}

\begin{appendix}
\section{Comparisons with Photoionization Models} \label{app:pyneb}
In this Appendix, we present photoionization models of the emissivities of several of the spectral features of importance in this work. We employ the \texttt{PyNeb} photoionization modeling code \citep{Luridiana2015}. \texttt{PyNeb} does not invoke a particular ionizing spectrum, instead modeling the emissivity of each species from an ionized gas regardless of the initial conditions of said gas. 

Given the relatively low spectral resolution of the G102 and G141 grisms, several of the emission lines are unresolved. We define the \neiii/\oii\ (Ne3O2) in terms of the coadded \oii\ doublet 
\begin{align}
  \frac{\neiii}{\oii}&\equiv\frac{\neiii~\lambda3869} {\oii~\lambda\lambda 3726,~3729} \label{eq:ne3o2}
\end{align}
and the \oiii/\oii\ ratio (O32) in terms of the coadded \oiii\ and \oii\ doublets
\begin{align}
    \mathrm{O}_{32}&\equiv\frac{\oiii~\lambda\lambda 4959,~5008 } {\oii~\lambda\lambda 3726,~3729} \label{eq:o32}
\end{align}

\begin{figure*}[t]
\epsscale{1.15}
\plotone{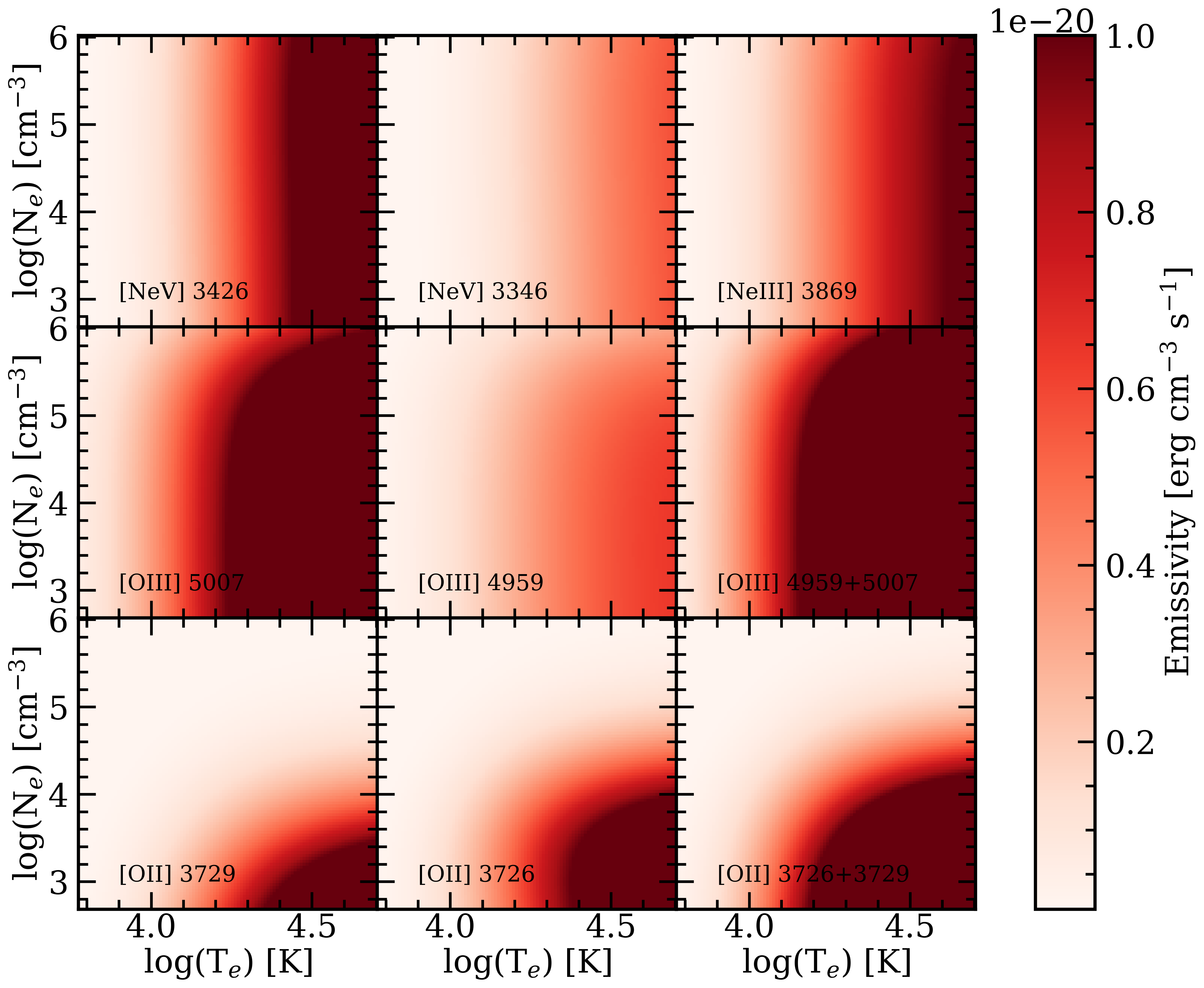}
\caption{\texttt{PyNeb} models for several relevant emission line emissivities as a function of temperature and density. The emissivities are all given on the same colormap scale. We see that the neon species shown here (\nev\ and \neiii) evolve only with temperature, while the oxygen species have evolution with both temperature and density in this parameter space. Given the spectral resolution of the \hst/WFC3 grisms, we also show the coadded \oii\ and \oiii\ emission grids. 
\label{fig:pyneb_models}}
\end{figure*}

In Figure \ref{fig:pyneb_models} we show emissivity maps as a function of temperature and density for several relevant emission features of neon and oxygen (\nev\ 3426 and 3346, \neiii\ 3869, \oiii\ 5007 and 4959, and \oii\ 3726 and 3729). We also show the coadded \oiii\ 5007+4959 and \oii\ 3726+3729 doublets. We note that the neon line emissivities do not significantly evolve with density and do evolve with temperature. The oxygen lines evolve both with temperature and density. 

\begin{figure*}[t]
\epsscale{1.15}
\plotone{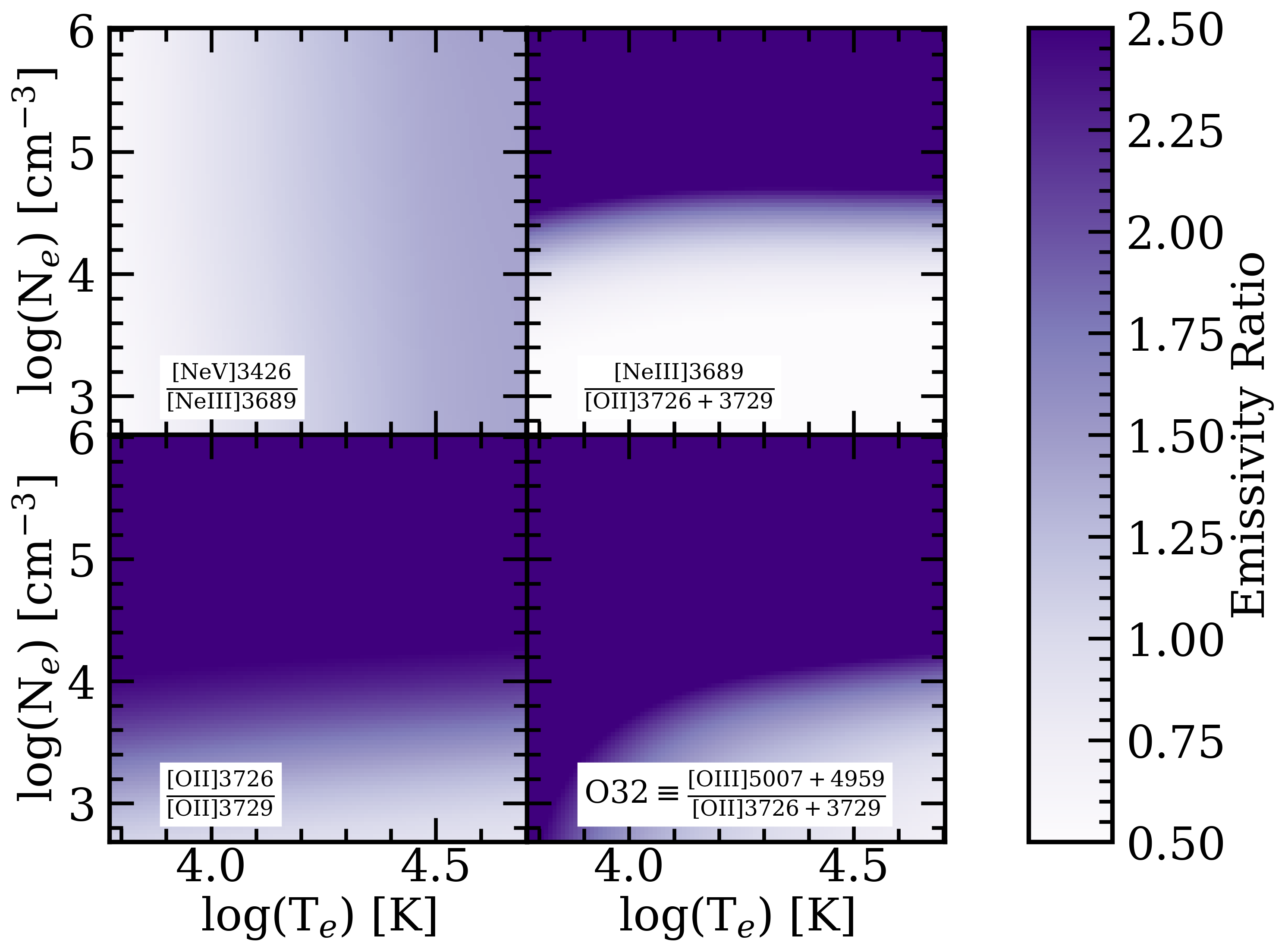}
\caption{\texttt{PyNeb} models for several relevant emission line ratios as a function of temperature and density. The ratio of the individual features of the \nev\ and \oiii\ doublets are constant in this temperature and density parameter space, and their constant ratios are given (2.73 and 2.98, respectively). The \nev/\neiii\ ratio is an indicator of temperature, where \oii/\oii\ and \neiii/\oii\ ratios are primarily functions of density.
\label{fig:pyneb_models_ratios}}
\end{figure*}

In Figure \ref{fig:pyneb_models_ratios} we show ratios of emissivities of several of the emission lines from Figure \ref{fig:pyneb_models} computed from \pyneb. The \nev\ 3426/\nev\ 3346 and \oiii 5007/\oiii 4959 ratios are both constant with temperature and density with values 2.73 and 2.98, respectively. We see that the \nev\ 3426/\neiii\ 3869 ratio increases with temperature. The \neiii/\oii\ ratio evolves primarily with density within this parameter space. 

We also show the \oii\ 3726/\oii\ 3729 ratio, which increases solely as a function of density in this parameter space. The \neiii/\oii\ and O32 ratios, as defined by equations \ref{eq:ne3o2} \ref{eq:o32}, respectively, vary with both temperature and density. 


\section{Sample Characteristics, Derived Quantities, and Emission-Line Fluxes} \label{app:data}
In this Appendix, we present the sample characteristics and derived quantities (Table \ref{tab:sample}) and the emission-line fluxes for relevant lines used in this work (Table \ref{tab:lines}).
\begin{deluxetable*}{llrrrrrr}
\tablecaption{Sample characteristics and derived quantities \label{tab:sample}}
\tablehead{
\colhead{Field} & \colhead{3D-HST ID} & \colhead{RA} & \colhead{Dec} & \colhead{Redshift\tablenotemark{i}} & \colhead{Stellar Mass\tablenotemark{ii}} & \colhead{$SFR_{UV}^{corr}$\tablenotemark{iii}} & \colhead{F435W - F775W\tablenotemark{iii}} \\ 
\colhead{} & \colhead{} & \colhead{Deg} & \colhead{Deg} & \colhead{} & \colhead{$\log(M_*/M_\odot)$} & \colhead{$M_\odot/$yr}  & \colhead{}
} 
\startdata
GN & 11743 & 189.2213 & 62.2002 & 2.0834 & 9.22 & 10.8 $\pm$0.9 & 0.26 $\pm$0.07 \\  
GN & 19464 & 189.0871 & 62.2376 & 2.0877 & 10.97 & 250 $\pm$10 & 1.0 $\pm$0.1 \\  
GN & 19591 & 189.2013 & 62.2380 & 2.002 & 10.25 & 260 $\pm$2 & 0.52 $\pm$0.03 \\  
GN & 19913 & 189.1483 & 62.2400 & 2.0104 & 11.57 & 1100 $\pm$40 & 0.84 $\pm$0.03 \\  
GN & 21290 & 189.2681 & 62.2462 & 2.2165 & 10.52 & 365 $\pm$3 & 0.05 $\pm$0.07 \\  
GN & 21412 & 189.1749 & 62.2476 & 2.0051 & 9.47 & 9.9 $\pm$0.2 & 0.51 $\pm$0.10 \\  
GN & 24427 & 189.3708 & 62.2617 & 1.7744 & 9.82 & 190 $\pm$10 & 0.30 $\pm$0.02 \\  
GS & 24803 & 53.1613 & -27.7958 & 1.4913 & 9.46 & 9.0 $\pm$0.7 & 0.49 \\  
GS & 26021 & 53.1604 & -27.7904 & 1.6117 & 9.52 & 12 $\pm$1 & 0.10\\  
GS & 28218 & 53.1469 & -27.7818 & 1.5516 & 9.05 & 4.0 $\pm$0.4 & 0.28\\  
GS & 29293 & 53.1583 & -27.7774 & 1.5523 & 8.87 & 3.89 $\pm$0.02 & -0.04\\  
GN & 37738 & 189.2812 & 62.3633 & 1.4506 & 11.07 & 106 $\pm$2 & 0.48 $\pm$0.02 \\  
GN & 37767 & 189.2960 & 62.3628 & 1.5181 & 9.08 & 7.2 $\pm$0.3 & 0.21 $\pm$0.09 \\  
GS & 38636 & 53.1269 & -27.7302 & 1.601 & 9.05 & 6.2 $\pm$0.7 & 0.05 \\  
GS & 40305 & 53.1228 & -27.7228 & 1.6132 & 10.68 & 70 $\pm$2 & 0.83 \\  
GS & 41195 & 53.1653 & -27.7185 & 1.9687 & 9.97 & \ldots & 0.31 \\  
GS & 41886 & 53.0977 & -27.7153 & 2.1397 & 10.87 & 183 $\pm$9 & 0.91   \\  
GS & 42614 & 53.0923 & -27.7122 & 1.6149 & 10.05 & 12 $\pm$2 & 0.34   \\  
GS & 42758 & 53.1122 & -27.7110 & 1.6085 & 10.38 & 1500 $\pm$900 & 0.71   \\  
GS & 44556 & 53.1081 & -27.7020 & 1.7736 & 9.34 & 6 $\pm$1 & 0.22   \\  
GS & 44783 & 53.1136 & -27.7014 & 1.6145 & 10.61 & 320 $\pm$10 & 0.78   \\  
GS & 45337 & 53.1117 & -27.699 & 1.6173 & 9.95 & 60 $\pm$1 & 0.19   \\
\enddata
\tablenotetext{i}{Grism redshifts from CLEAR \citep{Simons2023}}
\tablenotetext{ii}{From the 3D-HST catalog \citep{Skelton2014}. Masses have characteristic uncertainty of 0.3 dex.}
\tablenotetext{iii}{From the CANDELS/SHARDS catalog \citep{Barro2019}. Objects without uncertainties on photometry are nondetections in both filters.}
\end{deluxetable*}

\begin{deluxetable*}{llrrrrrrr}
\tablecaption{Observed Emission-Line Fluxes from CLEAR [$10^{-17}$erg s$^{-1}$cm$^{-2}$] \citep{Simons2023} \label{tab:lines}}
\tablehead{
\colhead{Field} & \colhead{ID} & \colhead{$\nev~\lambda3426$} & \colhead{$\oii~\lambda\lambda 3726,3729$} & \colhead{$~\lambda3869$} & \colhead{\Hb} & \colhead{$\oiii~\lambda\lambda4959,5007$} & \colhead{\Ha+$\nii~\lambda\lambda6548,6583$} & \colhead{$\sii~\lambda\lambda6717,6731$}
} 
\startdata
GN & 11743 & 2.83 $\pm$0.8 & 2.04 $\pm$0.78 & \ldots  & 0.69 $\pm$0.53 & 9.84 $\pm$0.6 & \ldots  &   \\  
GN & 19464 & 4.73 $\pm$1.37 & 4.54 $\pm$0.84 & 3.87 $\pm$0.95 & 1.19 $\pm$0.5 & 8.84 $\pm$0.55 &\ldots   & \ldots  \\  
GN & 19913 & 14.16 $\pm$1.92 & 4.81 $\pm$1.69 & 5.26 $\pm$3.16 & 8.56 $\pm$1.18 & 44.01 $\pm$1.41 & \ldots  & \ldots  \\  
GN & 21290 & 7.42 $\pm$0.87 & 7.11 $\pm$0.98 & 5.98 $\pm$1.3 & 4.91 $\pm$0.67 & 62.73 $\pm$0.96 &\ldots   & \ldots  \\  
GN & 21412 & 2.66 $\pm$0.63 & 2.1 $\pm$0.58 & 0.4 $\pm$0.95 & 1.24 $\pm$0.67 & 3.73 $\pm$0.74 &\ldots   &\ldots   \\  
GN & 24192 & 3.45 $\pm$0.53 & 3.12 $\pm$0.71 &\ldots   & 3.3 $\pm$0.48 & 15.55 $\pm$0.64 &\ldots   &\ldots   \\  
GN & 24427 & 10.76 $\pm$1.7 & 45.19 $\pm$1.43 & 5.42 $\pm$1.57 & 24.0 $\pm$1.05 & 140.77 $\pm$1.47 &\ldots   &\ldots   \\  
GS & 24803 & 2.48 $\pm$0.58 & 7.73 $\pm$0.4 & 1.21 $\pm$0.35 & 3.17 $\pm$0.49 & 15.3 $\pm$0.51 & 11.38 $\pm$0.39 & 2.45 $\pm$0.67 \\  
GS & 26021 & 1.57 $\pm$0.35 & 5.6 $\pm$0.24 & 0.98 $\pm$0.27 & 2.71 $\pm$0.36 & 8.14 $\pm$0.4 & 15.85 $\pm$4.08 &   \\  
GN & 28055 & 4.02 $\pm$0.84 &\ldots   &\ldots   & 4.04 $\pm$0.35 & 8.02 $\pm$0.38 &\ldots   &\ldots   \\  
GS & 28218 & 2.31 $\pm$0.72 & 2.2 $\pm$0.52 & \ldots  & \ldots  & 5.38 $\pm$0.6 & 4.73 $\pm$0.78 &\ldots   \\  
GS & 29293 & 2.17 $\pm$0.49 & 4.53 $\pm$0.33 & 0.97 $\pm$0.36 & 0.37 $\pm$0.42 & 13.21 $\pm$0.45 & 6.66 $\pm$0.65 &   \\  
GN & 37738 & 13.86 $\pm$1.36 & 22.77 $\pm$0.79 & 4.65 $\pm$0.69 & 8.13 $\pm$0.99 & 78.6 $\pm$1.38 & 47.53 $\pm$1.08 & 5.57 $\pm$1.0 \\  
GN & 37767 & 1.53 $\pm$0.48 & 1.51 $\pm$0.32 & 0.54 $\pm$0.29 & 3.23 $\pm$0.46 & 5.6 $\pm$0.53 &\ldots   &\ldots   \\  
GS & 38636 & 2.64 $\pm$0.69 & 4.34 $\pm$0.58 & \ldots  & 2.51 $\pm$0.74 & 8.78 $\pm$0.77 &\ldots   &\ldots   \\  
GS & 40305 & 3.61 $\pm$0.71 & 3.59 $\pm$0.51 & 0.98 $\pm$0.52 & 1.3 $\pm$0.98 & 8.58 $\pm$1.0 & 18.92 $\pm$8.11 &\ldots   \\  
GS & 41195 & 8.13 $\pm$1.12 & 29.22 $\pm$1.16 & \ldots  & 3.47 $\pm$2.01 & 25.76 $\pm$2.21 &\ldots   &\ldots   \\  
GS & 41218 & 2.14 $\pm$0.4 & 3.64 $\pm$0.87 & \ldots  & 1.53 $\pm$0.72 & 11.2 $\pm$0.85 &\ldots   &\ldots   \\  
GS & 41886 & 1.22 $\pm$0.4 & 2.29 $\pm$0.83 &\ldots   & 1.06 $\pm$0.53 & 14.46 $\pm$0.64 &\ldots   &\ldots   \\  
GS & 42614 & 5.71 $\pm$1.19 & 8.03 $\pm$0.82 &\ldots   &   & 8.09 $\pm$1.64 & 43.89 $\pm$8.87 &\ldots   \\  
GS & 42758 & 4.06 $\pm$0.91 & 14.43 $\pm$0.62 &\ldots   & 6.42 $\pm$0.83 & 14.73 $\pm$0.83 & 103.14 $\pm$10.23 &\ldots   \\  
GS & 44556 & 1.14 $\pm$0.36 & 3.04 $\pm$0.3 & 0.01 $\pm$0.31 & 1.79 $\pm$0.39 & 6.93 $\pm$0.46 &\ldots   &\ldots   \\  
GS & 44783 & 6.98 $\pm$1.41 & 26.31 $\pm$0.98 &\ldots   & 8.07 $\pm$1.72 & 19.59 $\pm$1.68 & 183.36 $\pm$20.2 &\ldots   \\  
GS & 45337 & 3.49 $\pm$1.15 & 28.53 $\pm$0.83 & 3.14 $\pm$1.13 &\ldots   & 34.76 $\pm$1.79 & 21.95 $\pm$7.72 &\ldots   \\  
\enddata
\end{deluxetable*}
\end{appendix}

\end{document}